\documentclass[structabstract]{aa}
\usepackage[varg]{txfonts}
\usepackage{natbib}
\usepackage{graphicx}
\usepackage{tabularx}
\usepackage{nicefrac}
\usepackage{units}
\usepackage{longtable}
\usepackage{longtable,lscape}

\bibpunct{(}{)}{;}{a}{}{,}
\newcommand{\referee}[1]{{\referee #1}}

\newcommand{\Eup} {$E{_{\rm up}}$}

\newcommand{\GHz}  {$\rm{GHz}$}

\newcommand{\bbf}{}
\newcommand{\new}{}
\begin{document}

%
%
\title{The HIFI spectral survey of AFGL~2591 (CHESS).\\ III. Chemical structure of the protostellar envelope.
\thanks{\textit{Herschel} is an ESA space observatory with science instruments provided by European-led Principal Investigator 
consortia and with important participation from NASA.}
}

\authorrunning{M. Ka{\'z}mierczak-Barthel et al.}
\titlerunning{Chemical structure of the AFGL~2591 envelope}
       
\author{M. Ka{\'z}mierczak-Barthel \inst{\ref{inst1}}
\and D.~A. Semenov \inst{\ref{inst2}}
\and F.~F.~S. van der Tak \inst{\ref{inst1},\ref{inst3}}
\and L. Chavarr\'{\i}a \inst{\ref{inst4}}
\and M.~H.~D. van der Wiel \inst{\ref{inst5}}
}

\institute{SRON Netherlands Institute for Space Research, Landleven 12, 9747 AD Groningen, The Netherlands\\
\email{maja.kazmierczak@gmail.com} \label{inst1}
\and Max Planck Institute for Astronomy, K{\"o}nigstuhl 17, D-69117 Heidelberg, Germany \label{inst2}
\and Kapteyn Astronomical Institute, University of Groningen, PO Box 800, 9700 AV, Groningen, The Netherlands \label{inst3}
\and Universidad de Chile, Camino del Observatorio 1515, Las Condes, Santiago, Chile \label{inst4}
\and Institute for Space Imaging Science, Department of Physics \& Astronomy, University of Lethbridge, Lethbridge AB, 
Canada \label{inst5}
}

\date{Submitted 23 July 2014 / Accepted 4 December 2014}

%
%
\abstract
  {}
   {The aim of this work is to understand the richness of chemical species observed in the isolated high-mass envelope of 
AFGL~2591, a prototypical object for studying massive star formation.}
   {Based on HIFI and JCMT data, the molecular abundances of species found in the protostellar envelope of AFGL~2591 were derived 
with a Monte Carlo radiative transfer code (\textsc{Ratran}), assuming a mixture of constant and 1D stepwise radial profiles for 
abundance distributions. The reconstructed 1D abundances  were compared with the results of the time-dependent gas-grain chemical 
modeling, using the best-fit 1D power-law {\bbf density} structure. The chemical simulations were performed considering ages 
of \mbox{1 $-$ 5 $\times\,10^4$~years,} cosmic ray ionization rates of \mbox{$5 - 500 \times 10^{-17}$~s$^{-1}$,} uniformly-sized 
$0.1-1~\mu$m dust grains, a dust/gas ratio of 1$\%$, and several sets of initial molecular abundances with C/O $<1$ and $>1$. The 
most important 
model parameters varied one by one in the simulations are age, cosmic ray ionization rate, external UV intensity, and grain size. 
}
   {Constant abundance models give good fits to the data for CO, CN, CS, HCO$^+$, H$_2$CO, N$_2$H$^+$, CCH, NO, OCS, OH, H$_2$CS, 
O, C, C$^+$, and CH.
Models with an abundance jump at 100\,K give good fits to the data for NH$_3$, SO, SO$_2$, H$_2$S, H$_2$O, HCl, and CH$_3$OH.
For HCN and HNC, the best models have an abundance jump at 230\,K.    
The time-dependent chemical model can accurately 
explain abundance profiles of 15 out of these 24 species. The jump-like radial profiles 
for key species like HCO$^+$, NH$_3$, and  H$_2$O are consistent with the outcome of the time-dependent chemical modeling. 
The best-fit model has a chemical age {\bbf of $\sim 10-50$~kyr}, a solar C/O ratio of $0.44$, and a cosmic-ray ionization rate 
of $\sim 5  \times 10^{-17}$~s$^{-1}$. The grain properties and the intensity of the external UV field do not strongly 
affect the chemical structure of the AFGL~2591 envelope, whereas its chemical age, the cosmic-ray ionization rate, and the initial 
abundances play an important role.}
{We demonstrate that simple constant or jump-like abundance profiles constrained 
with 1D \textsc{Ratran} line radiative transfer simulations are in agreement with time-dependent chemical modeling for most key C-, O-, N-, and S-bearing molecules. The main 
exceptions are species with very few observed transitions (C, O, C$^+$, and CH) or with a poorly established chemical 
network (HCl, H$_2$S) or whose chemistry is strongly affected by surface processes (CH$_3$OH).}

%
%

\keywords{{ISM: individual objects: AFGL 2591} - {Stars: formation} - {ISM: abundances: molecules: evolution} - {Submillimeter: 
ISM}}

\maketitle

\label{firstpage}

%
%
\section{Introduction} \label{section:introduction}
%
%
%
The processes that produce massive stars are complex and not very well understood. 
This is mainly because massive stars are relatively rare: about 1$\%$ of stars are more massive than 10~$M_{\sun}$. Their 
formation timescale is relatively fast and  shorter, by a factor of $\gtrsim$~10, than for low-mass stars 
\citep{2007ARA&A..45..481Z}. Moreover, massive stars remain deeply embedded for a considerable part of their 
lifetimes and usually form in clusters. While massive star-forming regions are typically several kiloparsecs away from our Sun, 
their challenging observations are the focus of many astrophysical investigations \citep[see,~e.g.,][]{2014arXiv1402.0919T}.

{\bbf This paper focuses on AFGL~2591,  a well-known, massive star-forming region with a circumstellar disk-like 
structure, bipolar outflows, and an extended envelope \citep{vanderTak1999}. It is a relatively isolated, deeply 
embedded, bright \citep[\mbox{$L = 2 \times\, 10^5\,L_{\sun}$},][]{Sanna2012} object. It is located at a distance of 
$3.33$\,kpc, which has  recently been accurately determined using trigonometric parallax measurements of H$_2$O masers 
\citep[][]{Rygl2012}.} 

{\bbf According to the modern paradigm of massive star formation, AFGL 2591 can be classified as a high-mass protostellar 
object or an early "hot core" \citep{2007A&A...475..549B,Veach2013}. The mass of the large-scale AFGL~2591 envelope 
is $\sim$400\,M$_{\sun}$ \citep{vanderTak2013}. Inside this envelope, a cluster of B-type stars is being formed, which reveals itself as 
several continuum emission sources detected at cm wavelengths with Very Large Array 
\citep[VLA~1, VLA~2, VLA~3, etc.][]{2003ApJ...589..386T,2005A&A...437..947V,2012ApJ...753...34J}. The derived masses of these 
young stellar objects are $\la 40\,M_{\sun}$ \citep{2012ApJ...753...34J}, and many of them are associated with prominent 
outflows (e.g., VLA~3). Among these sources, VLA~3 is one of the most massive objects ($\sim 38\,M_{\sun}$), which dominates the 
AFGL~2591 spectral energy distribution and luminosity \citep{2013A&A...551A..43J}.} 

{\bbf Using the Plateau de Bure interferometer, \citet{Wang2012} observed AFGL~2591 VLA~3 in dust continuum and lines of 
HDO, H$_2^{18}$O, and SO$_2$ at 0.5$^{''}$ resolution. Combining the data analysis with line radiative transfer calculations, they 
show that AFGL~2591 VLA~3 is surrounded by a massive disk with sub-Keplerian rotation and additional Hubble-like expansion 
motions. They also found that the observed molecular emission arises in various disk layers, with SO$_2$ tracing the disk atmosphere 
and HDO and H$_2^{18}$O probing the disk mid-plane. 
\citet{jimenezserra2012} used 0.5$^{''}$ resolution observations toward AFGL~2591 VLA~3 with the Submillimeter Array (SMA) and 
discovered its radial 
chemical differentiation down to linear scales $\la 3\,000$~AU. The derived excitation temperatures of $\sim 120-150$~K and the 
gas kinematics are suggestive of a Keplerian disk surrounding a massive, $40~M_{\sun}$ star.
%
} 



The physical structure of the circumstellar environment of AFGL~2591 was studied with single-dish submillimeter and infrared 
data as well as millimeter interferometry by \citet{vanderTak1999}. 
Based on a 1D model with a power-law density 
profile with index 1.25, these authors estimated the abundances of CS, HCO$^+$, HCN, and H$_2$CS. 
The dust temperature due to heating by 
the young star was calculated to exceed 120~K at $\sim$1500\,AU from the star, so that the ice mantles are expected to be 
fully evaporated in the inner envelope.

From modeling of C$^{17}$O emission lines, \citet{vanderTak2000} inferred that around 40$\% - $90$\%$ of the CO gas is 
depleted onto dust in the large-scale AFGL~2591 envelope. Their results 
suggest that freeze-out and thermal desorption control the gas-phase abundance of CO. \citet{vanderTak2003} 
observed sulfur-bearing molecules in several massive protostellar envelopes, and derived column densities and 
abundances of H$_2$S, 
SO, SO$_2$, H$_2$CS, HCS$^+$, NS, OCS, and CS in AFGL~2591. These results are consistent with a model of ice evaporation in the 
envelope with gradients in temperature and density, and a chemical age of $\sim 10^5$ years.

\citet{Bruderer2010b} used the HIFI instrument onboard the \textit{Herschel} satellite and  detected emission lines of light 
hydrides such 
as CH, CH$^+$, NH, OH$^+$, and H$_2$O$^+$. They found that observed abundances and line excitation of CH and CH$^+$ can be 
explained with an envelope-outflow model, where far-ultraviolet (FUV) stellar radiation irradiates and heats the outflow walls, 
driving the gas-phase, \mbox{photon-dominated} chemistry that produces these molecules.


{\bbf Using these large observational data sets of the AFGL~2591 molecular content, a number of theoretical studies 
of its chemical and physical evolution have been performed.}
The study of the envelope AFGL~2591 by \citet{Doty2002} was based on a 1D thermal and gas-phase chemical model of the 
observed abundances and column densities for 29~species. They were able to explain most of the chemical structures and 
found hints of high-temperature nitrogen and hydrocarbon chemistry. They also derived that most of the sulfur in the cold 
gas should be locked in CS, in SO$_2$ in the warm gas, and S in the hot gas. From the observed abundances of HCO$^+$ 
and N$_2$H$^+$ and the cold gas-phase production of HCN the cosmic-ray ionization rate was constrained, $\sim 
5\times10^{-17}$~s$^{-1}$. The derived best-fit chemical age is between $7\,000$ and $50\,000$~years.

\citet{Stauber2004,Stauber2005} analyzed the influence of FUV and X-ray radiation from massive young stellar objects 
on the chemistry of its envelope. The UV flux seems to affect only the regions within 500-600\,AU from the star, but X-rays can 
penetrate larger distances. They both enhance the abundance of simple hydrides, ions, and radicals.

\citet{Bruderer2009b,Bruderer2009,Bruderer2010} introduced a 2D model with a self-consistent calculation of the dust 
temperature, a fast chemical method, and a multi-zone escape probability method for the line radiative transfer. They benchmarked 
the feasibility of this model by comparing it with published models of the AFGL~2591 protostellar envelope. \citet{Bruderer2010b} 
utilized this 2D model to investigate the effects of the cavity shapes, outflow densities, and disk properties on the large-scale 
chemical composition of AFGL~2591. They derived that the \textit{Herschel} observations of diatomic hydrides and hydride ions can 
be explained by the outflow model with the walls directly irradiated by the stellar FUV photons, producing high gas temperatures 
and rapid photodissociation of their saturated parent species.

We observed AFGL~2591 with the Heterodyne Instrument for the Far-Infrared \citep[HIFI,][]{Graauw2010} on board the ESA 
\textit{Herschel Space Observatory} \citep{Pilbratt2010} as a part of the HIFI/CHESS Guaranteed Time Key Programme \citep[Chemical 
Herschel Survey of Star Forming Regions,][]{Ceccarelli2010}. A full spectral survey of AFGL~2591 of HIFI bands 
1a--5a (480--1240~\GHz) was obtained. Nine additional selected frequencies were observed; the corresponding 
bands are 5b (lines: HCl and CO), 6a (CO), 6b (CO), 7a (NH$_3$, CO), and 7b (CO, OH, [CII]). 
These observations were supplemented with several lines from the ground-based James Clerk Maxwell Telescope 
(JCMT)\footnote{The James Clerk Maxwell Telescope is operated by the Joint Astronomy Centre on behalf of the Science and 
Technology Facilities Council of the United Kingdom, the Netherlands Organisation for Scientific Research, and the National 
Research Council of Canada.} as part of the JCMT Spectral Legacy Survey \citep[SLS,][]{Plume2007,vanderWiel2011}.

The results from our spectral survey are presented in a series of {\new three} papers. The first paper focused on highly-excited linear  
molecules \citep[][{\bbf Paper~I}]{vanderWiel2013}. We analyzed the lines profiles of CO isotopologues, HCO$^+$, CS, HCN, and 
HNC in the FIR/mm range by using a 1D non-LTE line radiative transfer model. It was found that a spherical envelope model is able to 
fit the upper-state transitions at $E_{\rm ul}/k > 150$~K for all species, and that an additional hot, optically thin gas 
component (such as from an outflow cavity wall) is required. In the second paper the entire survey was summarized 
\citep[][{\bbf Paper~II}]{2014arXiv1405.4761K}. A simple excitation analysis was also performed to estimate column densities and 
excitation temperatures for a~multitude of observed molecules. In total, we identified 32 species (including isotopologues), with 
268 emission and 16 absorption lines (excluding blended lines). The derived molecular column densities range from $6\times 
10^{11}$~cm$^{-2}$ to $10^{19}$~cm$^{-2}$ {\bbf for N$_2$H$^+$ and CO}, and excitation temperatures range from 19 to 175~K {\bbf for HCO$^+$ and 
SO$_2$, respectively}. We have distinguished two gas components, namely cold gas traced by e.g., HCN, H$_2$S, and NH$_3$ ($T\la 
70$~K) and warm gas traced by e.g., CH$_3$OH and SO$_2$ ($T >70-100$~K).

In the present work {\bbf (Paper III)}, we are going one step further in the analysis of our spectral survey. Here, all 
molecules observed in the protostellar envelope are modeled using the radiative transfer code \textsc{ratran} 
\citep{Hogerheijde2000} together with a time-dependent gas-grain chemical model based on the \textsc{alchemic} code 
\citep{Semenov_ea10}. 

This paper is organized as follows. Section 2 presents a~radiative transfer analysis of molecules observed in the protostellar 
envelope of AFGL~2591, based on all data listed in Paper~II, and Sect. 3 describes the chemical modeling. 
Discussion of our results and comparison between these two approaches is given in Sect. 4, and our conclusions follow in Sect. 5.

%
%
\section{Radiative transfer modeling} \label{Ratran}
%
%
%
%
\subsection{One-dimensional physical model} \label{sec:phys_model}
\begin{figure}
\centering
\includegraphics[width=0.24\textwidth]{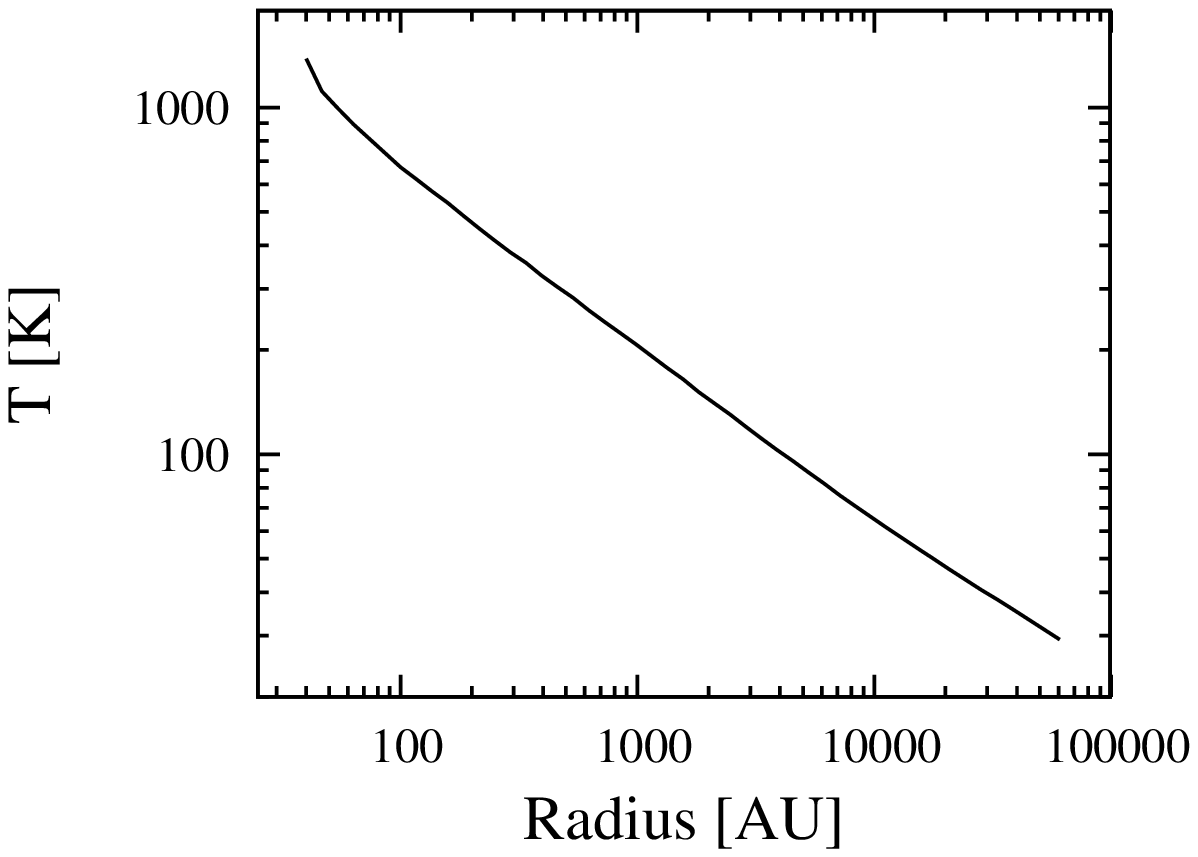}
\includegraphics[width=0.24\textwidth]{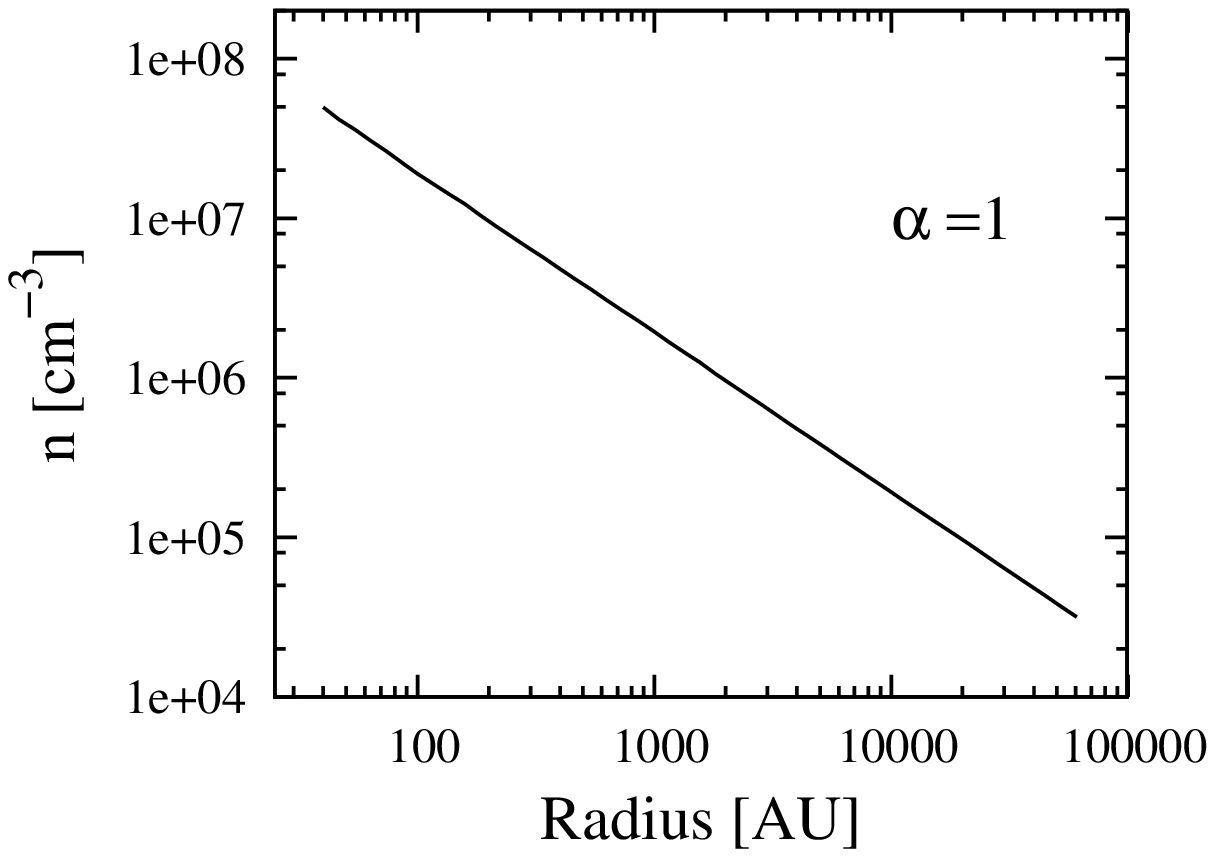}
\caption{The temperature and density structure of the 1D physical model of AFGL~2591.}
\label{figure:afgl2591_profile}
\end{figure}

%
%
The model for the physical structure of the envelope of AFGL~2591 is based on the study of \citet{vanderTak2013}. It~uses the new 
distance of 3.3\,kpc and the corresponding luminosity of \mbox{$L = 2 \times\,10^5$\,$L_{\sun}$.} The previous distance estimates 
were uncertain, with values between 1 and 2\,kpc \citep[e.g.,][]{vanderTak1999,vanderTak2000}, thus the luminosity at 1\,kpc is 
\mbox{$L = 2 \times\,10^4\,L_{\sun}$.} The radial density profile was derived from the JCMT/SCUBA\footnote{SCUBA is a 
submillimeter continuum array receiver operated on the James Clerk Maxwell Telescope, on Mauna Kea, Hawaii.} (450, 850\,$\mu$m) 
and the \textit{Herschel}/PACS\footnote{PACS \citep{Poglitsch2010} is the Photodetector Array Camera and Spectrometer onboard the 
ESA \textit{Herschel} Space Observatory.} (70, 100, 150\,$\mu$m) data with the radial density power-law index of $\alpha \sim 1$, 
assuming a power-law density profile \mbox{$n(r) \propto r^{-\alpha}$.} The physical model consists of 50 shells, with 
temperatures of 1383\,K and 29\,K in the inner and outer shells, respectively, and a total gas mass of 373\,$M_{\sun}$. The 
temperature and density structure of the model are presented in Fig.~\ref{figure:afgl2591_profile}.

The results   presented in this paper are based on a~slightly different physical model of AFGL~2591 from that described in Paper~I 
\citep{vanderWiel2013}. The model from Paper~I was based only on the SCUBA data; here the PACS data were added \citep[for the 
modeling procedure see][]{vanderTak2013}, and so all molecular abundances from the Papers I and III were recalculated 
with the new physical model that is based on the combined SCUBA and PACS data sets.

%
%
\subsection{Line modeling} \label{sec:1D}
%
%

Spherically symmetric radiative transfer calculations with the Monte Carlo \textsc{ratran} code \citep{Hogerheijde2000} were run 
to estimate abundances of observed species in the protostellar envelope of AFGL~2591. Spectroscopic and collisional parameters 
were taken from the \textsc{lamda}\footnote{\tt http://home.strw.leidenuniv.nl/$\sim$moldata/} database \citep[Leiden Atomic and Molecular Database,][]{lamda}. 
Detailed information about the adopted collisional rate coefficients for each molecule is given in Sect.~\ref{Ratran-molecules}. 

As a first step, abundances were estimated assuming a constant relative abundance throughout the protostellar envelope. To find 
the best-fit values of the abundances, observed and synthetic line fluxes were compared using $\chi^2$-minimization. The 
constant-abundance models agree with the data of several molecules in terms of the total line flux and the shape of the line. 

However, modeling the observed lines with a constant abundance may not always be the correct approach. Laboratory studies indicate 
that many molecules reside in more or less volatile ice mantles on dust grain surfaces at temperatures  
below $20 - 110$\,K depending on species, surface composition, and structure (see, e.g., \citet{1998ARA&A..36..317V}, 
\citet{2006A&A...449.1297B}, \citet{2009ARA&A..47..427H}, and \citet{2011A&A...529A..74F}). Usually, water ice is a major solid 
component that entraps other heavy ices and a fraction of volatile ices such as CO. Hence, models with jump-like abundance 
profiles at $\sim$100\,K may better represent the observations of gas-phase species in warm environments. 

\citet{2002A&A...390.1001S} performed detailed 1D radiative transfer analysis of the observed continuum and lines toward the young 
stellar object IRAS~16293-2422. They concluded that the lines of some molecules, e.g., H$_2$CO, CH$_3$OH, SO, SO$_2$, and OCS can 
only be explained when their abundances are greatly increased in warm regions of the envelope, where ices are expected to thermally 
evaporate. \citet{2004A&A...416..603J} derived that in low-mass pre- and protostellar objects, the spectra of nitrogen-bearing 
species together with HCO$^+$ and CO cannot be fitted with a constant abundance and require a model with a radial chemical gradient, 
where the region with molecular freeze-out is represented by a jump-like drop of the molecular abundances. \citet{vanderTak2006} 
studied water line spectra in the envelopes and disks around young massive stars and found that the water abundance is lower 
outside a critical radius, which corresponds to a temperature of $\sim 100$~K.

According to the chemical study of \citet{Rodgers2001}, the temperature of 230~K is also important, and may cause another sudden, 
jump-like rise in the abundances of N-bearing species. Their chemical model predicts that high-temperature neutral-neutral reactions 
of O and OH with H$_2$ drive the remaining oxygen that is not bound in CO into H$_2$O. This partly breaks the chemical cycling 
between N$_2$ and N, where N$_2$ (which is constantly destroyed by ionized He atoms) is reformed by rapid reactions of released N 
with OH first into NO and NH, and then back to N$_2$. As a result, at $T\ga 230$~K abundances of major N-bearing molecules become 
greatly enhanced. \citet{Boonman2001} have analyzed the HCN $J$=9--8 spectrum from the AFGL~2591 envelope 
and found that the HCN abundance above the temperature of 230~K should be at least 2~orders of magnitude higher than in the colder 
region.

Thus, in the present study of the protostellar envelope of AFGL~2591, the observations of each molecule are analyzed in the 
framework of the constant abundance model as well as models with jumps at 100\,K and 230\,K. The molecular abundances were estimated 
from the analysis of the optically thin isotopologue lines if their observations were available. We used the standard isotopic 
ratios: \mbox{$^{12}$C/$^{13}$C = 60,} \mbox{$^{16}$O/$^{18}$O = 500,} \mbox{$^{16}$O/$^{17}$O = 2500,} \mbox{$^{14}$N/$^{15}$N = 
270,} and \mbox{$^{32}$S/$^{34}$S = 22} \citep{Wilson1994}.

%
%
\subsection{Error budget} \label{Ratran-errors}
%
%
%
\subsubsection{Molecular data} \label{Ratran-errors-chem}
%
%
One of the largest sources of uncertainty for radiative transfer analysis are molecular data, especially collisional rate 
coefficients. \citet{lamda} and \citet{vanderTak2011IAUS280449V} describe in detail the accuracy of molecular data needed for 
radiative transfer analysis, such as collisional rate coefficients, spontaneous emission probabilities, transition frequencies, 
statistical weights, and energy levels. The uncertainties involved in determining collisional rate coefficients are summarized by 
\citet{Flower1990mcimbook}, \citet{Roueff1990moasbook}, and \citet{2013ChRv1138906R}.
The determination of collisional rate coefficients includes uncertainties that vary for different species. First, there is an 
inconsistency in the astrophysical literature regarding choices of values of electric dipole moments for some molecules, which may cause 
an error of a factor of two in their derived abundances \citep{lamda}. Second, published collisional rate coefficients often cover 
only a limited range of temperatures and energy levels, and extrapolation of their values outside of these ranges may cause 
additional errors \citep{vanderTak2011IAUS280449V}. Third, many molecules do not have calculated collisional rate coefficients at all or 
have the collisional data available only for collisions with $e^-$, He, or \mbox{H$_2$ $J = 0$} \citep{lamda}. For some molecules 
it is possible to scale rates obtained for a molecule with similar structure, using the reduced mass \citep[see Eq.~(12),][]{lamda}. 
Such scaling can be dangerous when collisional rates with helium are used as a proxy for the H$_2$ collisional rates. In addition, 
for freshly formed or hot H$_2$ gas collisions with \mbox{H$_2$, $J = 1$} can become important to properly describe the level 
populations, which is often ignored in astrophysical modeling. All together, these uncertainties may cause the estimated 
abundances to have error bars of $20\%$ and probably higher.  

%
\subsubsection{Physical model} \label{Ratran-errors-phys}
%
%
The uncertainties associated with the  physical model of AFGL~2591 mostly depend on the estimation of the envelope mass, density 
profile, and distance to the object \citep{vanderTak2013}. The distance to AFGL~2591, \mbox{$d = 3.33 \pm 0.11$~kpc,} has been quite 
 accurately determined thanks to the VLBI parallax measurements of 22\,GHz water maser emission \citep[see][]{Rygl2012}. 
The influence of the envelope masses and density profiles on the physical model were analyzed in the same way as described in 
Sect.~5.5 of \citet[][]{Herpin2012A&A542A76H}. In our model, the main uncertainties arise from the adopted dust opacities. Hence, we 
performed test SED modeling taking these uncertainties into account and found that the resulting CO abundances in the AFGL~2591 
envelope can differ by a factor of $\sim 3$, which is the dominant factor. {\bbf However, this parameter affects all our abundance estimates 
equally, so that the relative abundances between species are unchanged.}

%
\subsubsection{Observational data} \label{Ratran-errors-obs}
%
%
Regarding observational uncertainties, we have to take into account errors from the measurements of the integrated line intensities, 
because these values were iteratively compared with the modeled line fluxes to identify a best-fit model. For strong lines (e.g., 
CO), calibration errors dominate the uncertainty. For Herschel data this is $\sim 10\%$; for the ground-based data the calibration 
uncertainties are more than $20\%$. For weak lines, spectral noise dominates the error, which amounts to \mbox{50 $-$ 100\,mK} 
depending on frequency. All in all, the total error is the maximum of these contributions.

The other factor that affects the accuracy of estimated abundances is the number of observed transitions for a given molecule. The 
abundances of a species can be calculated more precisely when observations of different lines from many energy levels are 
available, such as many different transitions of CO and its isotopologues. On the other hand,  for NO, OCS, OH, H$_2$CS, C, C$^+$, 
CH, and O, observations of only one or two lines are available. Consequently, the uncertainties of their best-fit abundances are 
higher.

%
\subsection{Derived abundances of analyzed species} \label{Ratran-molecules}
%
%
In Table~\ref{table:Ratran-abundances} the best-fitting values of abundances for observed species in the protostellar envelope of 
AFGL~2591 are presented. The table is divided into five parts, corresponding to the five following subsections in which we discuss   the 
modeled species. 

In the first part of Table~\ref{table:Ratran-abundances} (Sect.~\ref{cons1}) molecules (CO, CN, and CS) for which a constant model 
gave the best-fitted abundance are listed. Moreover, for these species we observed  many transitions from different 
energy levels, so it was possible to exclude models with jumps at 100\,K and 230\,K. 

The second part of Table~\ref{table:Ratran-abundances} (Sect.~\ref{cons1b}) consists of species (HCO$^+$, H$_2$CO, N$_2$H$^+$, and 
CCH) for which a constant model gave as good a fit to the observed fluxes as a model with a jump at 100~K, with a lower abundance in 
the inner part of the envelope. For these four molecules, the best-fit constant model gives the same abundance as is derived in 
the outer part of the envelope from the model with a jump. The lower inner abundance was predicted by chemical models and then 
applied to \textsc{ratran}. Both models, constant and with a jump, reproduce very well  our observed integrated line intensities. We 
also observed several transitions from different energy levels of these molecules, so it was possible to exclude jump models with a
higher inner abundance. 

Part 3 of Table 1 presents species (NO, OCS, OH, H$_2$CS, C, C$^+$, CH, and O) for which a constant model also gave the best-fit 
abundance; however, it is not possible to exclude a jump model because of the lack of different observed transitions. Usually, for 
these species we observed only one line or a few lines (Sect.~\ref{cons2}), which causes higher uncertainties in the derived abundances 
of these species, up to a factor of 12 for CH and C. 

Only two molecules are listed in the fourth part of Table~\ref{table:Ratran-abundances} (Sect.~\ref{jump230}), HCN and HNC. The 
model with an abundance jump at 230\,K was found to provide the best-fit value of $\chi^2$. This is consistent with the results of 
\citet{Boonman2001} for HCN. 

In the last part of Table~\ref{table:Ratran-abundances} (Sect.~\ref{jump100}) we list the~molecules (NH$_3$, SO, SO$_2$, H$_2$S, 
H$_2$O, HCl, and CH$_3$OH) for which the model with jump at 100\,K provided the best fit. The~constant models were run first for 
these molecules, but they either did not reproduce the observed line fluxes (SO, SO$_2$, and CH$_3$OH) or made some lines  
appear in absorption (NH$_3$, H$_2$S, H$_2$O, and HCl).

\begin{table}
\setlength{\extrarowheight}{2pt}
\caption{Abundances {\new (relative to H$_2$)} estimated with the spherical radiative transfer model (\textsc{ratran}) for the envelope of AFGL 2591, based 
on HIFI and JCMT data.}
\label{table:Ratran-abundances}
\centering
\begin{tabular}{c|ccc}
\hline\hline
Molecule & Inner abundance & Outer abundance & Jump at\\ 
\hline
CO        & \multicolumn{2}{c}{2 $\times$ 10$^{-4}$} & const\\ 
CN        & \multicolumn{2}{c}{1 $\times$ 10$^{-9}$}& const\\ 
CS        & \multicolumn{2}{c}{4 $\times$ 10$^{-8}$} & const\\ 
\hline
HCO$^+$   & 3 $\times$ 10$^{-11}$ & 3 $\times$ 10$^{-8}$& 100\,K\\ 
H$_2$CO   & 1 $\times$ 10$^{-11}$ & 1 $\times$ 10$^{-8}$& 100\,K\\ 
N$_2$H$^+$& 8 $\times$ 10$^{-13}$ & 8 $\times$ 10$^{-10}$& 100\,K\\ 
CCH       & 8 $\times$ 10$^{-12}$ & 8 $\times$ 10$^{-8}$& 100\,K\\ 
\hline
NO$^*$     & \multicolumn{2}{c}{5 $\times$ 10$^{-8}$} & const\\ 
OCS$^*$    & \multicolumn{2}{c}{4 $\times$ 10$^{-8}$} & const\\ 
OH$^*$     & \multicolumn{2}{c}{2 $\times$ 10$^{-8}$} & const\\
H$_2$CS$^*$& \multicolumn{2}{c}{4 $\times$ 10$^{-9}$} & const\\ 
C$^*$     & \multicolumn{2}{c}{4 $\times$ 10$^{-6}$} & const\\ 
C$^+$$^*$    & \multicolumn{2}{c}{7 $\times$ 10$^{-6}$} & const\\ 
CH$^*$     & \multicolumn{2}{c}{1 $\times$ 10$^{-8}$} & const\\ 
O$^{**}$  & \multicolumn{2}{c}{2 $\times$ 10$^{-5}$} & const\\ 
\hline
HCN       & 5 $\times$ 10$^{-6}$ & 5 $\times$ 10$^{-7}$ & 230\,K\\ 
HNC       & 1 $\times$ 10$^{-7}$ & 1 $\times$ 10$^{-8}$ & 230\,K\\ 
\hline
NH$_3$    & 4 $\times$ 10$^{-6}$ & 4 $\times$ 10$^{-10}$& 100\,K\\ 
SO        & 5 $\times$ 10$^{-7}$ & 5 $\times$ 10$^{-9}$ & 100\,K\\ 
SO$_2$    & 5 $\times$ 10$^{-6}$ & 5 $\times$ 10$^{-9}$ & 100\,K\\ 
H$_2$S    & 8 $\times$ 10$^{-8}$ & 8 $\times$ 10$^{-10}$ & 100\,K\\ 
H$_2$O    & 1 $\times$ 10$^{-4}$ & 1 $\times$ 10$^{-9}$ & 100\,K\\ 
HCl       & 4 $\times$ 10$^{-6}$ & 4 $\times$ 10$^{-10}$ & 100\,K\\ 
CH$_3$OH  & 8 $\times$ 10$^{-7}$ & 8 $\times$ 10$^{-8}$ & 100\,K\\ 
\hline
\end{tabular}
  \begin{list}{}{}
        \item[* ~-- ] indicates highly uncertain abundances
        \item[** ~-- ] based on PACS data (W. Kwon, priv. comm.)
   \end{list} 
\end{table}

%
\subsubsection{Constant abundance model with many transitions} \label{cons1}
%
%
In this section we discuss species (CO, CN, and CS) modeled by a constant abundance model (first part of 
Table~\ref{table:Ratran-abundances}). For these species we observed many transitions from different energy levels, and models with 
jumps at 100\,K and 230\,K were excluded.

Carbon monoxide is one of the most widely studied and well-known molecules. Because of its importance as a molecular tracer, detailed and 
accurate CO molecular data are available. For this work, we adopted the collisional rate coefficients for CO {\bbf with H$_2$} 
from \citet{Yang2010}, which cover energy levels up to $J=40$ for temperatures ranging from 2\,K to 3\,000\,K. We estimate a 
CO abundance relative to H$_2$ of \mbox{2 $\times$ 10$^{-4}$}, which is the same value as previously calculated by  
\citet{vanderTak1999}, although    the two results are based on different physical models. Our result was based on 
observations of 31 lines of CO and its isotopologues ($^{13}$CO, C$^{18}$O, and C$^{17}$O); however, taking into account the 
possible sources of uncertainties, the overall CO abundance uncertainty was estimated to be a factor of~3.

Collisional rates for CN {\bbf with He} were calculated by \citet{Lique2010}; {\bbf the scaled rates for collisions with 
H$_2$ were adopted from the \textsc{lamda} database}. We obtained an abundance of \mbox{1 $\times$ 10$^{-9}$} with the uncertainty 
within a factor of~5. \citet{Stauber2007} estimated a higher CN abundance of \mbox{2.3 $\times$ 10$^{-8}$} by using a different 
physical model and assuming a distance of \mbox{$d = 1$\,kpc}. These authors also tested models with a jump in abundance at 
100\,K and estimated an outer abundance of \mbox{10$^{-9}$} and an inner abundance of \mbox{6 $\times$ 10$^{-7}$.} However, the two 
models were statistically indistinguishable. \citet{Stauber2007} based their calculations on seven~hyper-fine components from 
\mbox{\Eup $\sim$ 33\,K}. Our estimations of the CN abundance are based on several CN lines from three different energy levels of 
\mbox{\Eup\, $\sim$ 33, 82, and 114\,K} and are thus more robust. The overall CN abundance uncertainty is estimated to be \mbox{a 
factor of~3.}

We scaled the rate coefficients for collisions of CS and H$_2$ from collisions of CS with He, which were taken from 
\citet{Lique2006}. Based on observations of several lines of CS from  energy levels of \mbox{\Eup\,$ = 66 - 282$\,K} and 
its two isotopologues, $^{13}$CS and C$^{34}$S, we derived a CS abundance of \mbox{4 $\times$ 10$^{-8}$}. The uncertainty of this 
estimation is within a factor of~4. \citet{vanderTak2000} estimated a slightly lower abundance, \mbox{[CS] = 1 $\times$ 
10$^{-8}$}, but using the old distance and an outdated model.

%
\subsubsection{Model with a lower inner abundance} \label{cons1b}
%
%

In this section we discuss species (HCO$^+$, H$_2$CO, N$_2$H$^+$, and CCH) fitted by a model with an abundance jump at 100\,K, 
with a lower inner abundance (second part of Table~\ref{table:Ratran-abundances}). This model gives an equally good fit to the 
observational fluxes as a constant abundance model. The lower abundance in the inner part of the envelope was predicted by 
chemical models (Sect. \ref{sec:chem_model}).

Collisional rates for HCO$^+$ {\bbf with H$_2$} were taken from \citet{Flower1999}. From \textsc{ratran} calculations we 
derived, with the uncertainty within a factor of 4, the following HCO$^+$ abundances: \mbox{[HCO$^+$] = 3 $\times$ 10$^{-11}$} in 
the inner part of the envelope \mbox{(where $T >$ 100\,K)} and \mbox{[HCO$^+$] = 3 $\times$ 10$^{-8}$} in the outer region of the 
envelope \mbox{($T <$ 100\,K).} The outer abundance is very similar to the value calculated  by van der Tak et al (1999): \mbox{1 
$\times$ 10$^{-8}$}.

The collisional rates for H$_2$CO {\bbf with H$_2$} were taken from \citet{Wiesenfeld2013}. Based on 17 H$_2$CO transitions 
from a broad range of energy levels (\Eup\,= 32 $-$ 236\,K), we derived an abundance of \mbox{1 $\times$ 10$^{-11}$} and \mbox{1 
$\times$ 10$^{-8}$} in the inner and outer envelope with the uncertainty of a factor of~4. The ortho-to-para ratio is 3:1 for 
H$_2$CO, thus models for o-H$_2$CO and p-H$_2$CO were run separately and then  both abundances were added to get a total 
H$_2$CO abundance. Based on the old distance and physical model, \citet{vanderTak1999,vanderTak2000} estimated the H$_2$CO 
abundance in the range of \mbox{$2 - 4$ $\times$ 10$^{-9}$} which is between our values.

The collisional rate coefficients for N$_2$H$^+$ {\bbf with H$_2$ were adopted from the \textsc{lamda} database, where they 
were scaled from the rates for HCO$^+$ from \citet{Flower1999}.}
Based on three observed lines, we obtained N$_2$H$^+$ abundances of \mbox{8 $\times$ 10$^{-13}$} and \mbox{8 $\times$ 
10$^{-10}$} in the inner and outer part of the envelope, respectively, with an uncertainty of a factor of~5.

The rate coefficients for collisions of CCH with H$_2$ were scaled from collisions of CCH with He, which were adopted from 
\citet{Spielfiedel2012}. We used eight lines from four different energy levels and derived an abundance of \mbox{8 $\times$ 10$^{-12}$} 
and \mbox{8 $\times$ 10$^{-8}$} in the inner and outer parts of the envelope. The error of this estimation is a factor of~4.

%
\subsubsection{Constant abundance model with a few transitions} \label{cons2}
%
%
In this section we discuss species (NO, OCS, OH, H$_2$CS, C, C$^+$, CH, and O) that were modeled by a constant abundance model 
(third part of Table~\ref{table:Ratran-abundances}). For these species we observed only one or two transitions from different 
energy levels. Thus, the precision of the estimated abundances is lower and models with a jump at 100\,K or 230\,K cannot be 
excluded.

{\bbf The collisional rate coefficients for NO with H$_2$ were adopted from the \textsc{lamda} database, where they were 
scaled from the rates for NO with He by \citet{Lique2009}.}
We obtained abundances of \mbox{[NO] = 5 $\times$ 10$^{-8}$} and a very similar value was estimated by \citet{Stauber2007}, 
\mbox{[NO] = 3.1 $\times$ 10$^{-8}$.} However, our calculations are based on only four lines from two different energy levels, which 
causes  uncertainties that are higher by a factor of $\sim$8. We  also used a different physical model than \citet{Stauber2007}.

The collisional rates for OCS were adopted from \citet{Green1978ApJS7169G}. \citet{Flower2001MNRAS328147F} calculated the rate 
coefficients for collisions of OCS \mbox{(J $\le$ 29)} with He at low temperatures \mbox{(T $\le$ 150\,K)} and found that their 
results are comparable with those for collisions between OCS and p-H$_2$ derived by \citet{Green1978ApJS7169G}. We derived an 
abundance of \mbox{4 $\times$ 10$^{-8}$}, which is a very similar value to that previously calculated  by \citet{vanderTak2003}, 
\mbox{1 $\times$ 10$^{-8}$.} However, we also consider this result to be very uncertain (a factor of~8) because this estimation is 
based on only three lines from very similar energy levels.

The collisional rate coefficients for OH {\bbf with H$_2$} were taken from \citet{Offer1994JChPh100362O}. We derived 
\mbox{[OH] = 2 $\times$ 10$^{-8}$}, but this estimation is very uncertain (a factor of~8), because it is based on fitting the 
observations of only one transition.

We scaled the collisional rate coefficients for H$_2$CS from the rates for H$_2$CO, which were taken from \citet{Wiesenfeld2013}. 
We derived \mbox{[H$_2$CS] = 4 $\times$ 10$^{-9}$.} This result is also uncertain by a factor of~8 because it is based on only~three 
lines from very similar energy levels. All three observed lines belong to \mbox{o-H$_2$CS.} The total H$_2$CS abundance was 
calculated assuming an ortho-to-para ratio of 3:1. \citet{vanderTak2003} derived a lower H$_2$CS abundance, \mbox{3 $\times$ 
10$^{-10}$}, using older molecular data, and an old physical model using the old distance. 

Rate coefficients for collisions of C atoms with H$_2$ were taken from \citet{Schroder1991JPhB242487S}. Based on two 
observed transitions, we derived \mbox{[C] = 4 $\times$ 10$^{-6}$} with an error of a factor of $\sim$8.

Rate coefficients for collisions for C$^+$ with H$_2$ were adopted from \citet{Flower1977JPhB103673F}. 
We derive a C$^+$ abundance of \mbox{7 $\times$ 10$^{-6}$.} This result is highly uncertain (a factor of $\sim$12), again because it is 
based only on one observed transition. Moreover, the C$^+$ line profile is contaminated by emission from the off-position, even after applying 
corrections, which causes a higher error of the measurements {\bbf (see Appendix of Paper II)}.

Currently, accurate collisional rates for CH do not exist, so we scaled OH collisional rate coefficients \citep{Offer1994JChPh100362O}. 
This scaling may cause significant errors, in particular because CH and OH do not have the same electronic structure.
Hence we consider our abundance estimate of \mbox{1 $\times$ 10$^{-8}$} as highly uncertain (a factor of $\sim$12).

Rate coefficients for collisions for O with H$_2$ were taken from \citet{1992JPhB25285J}. The derived O abundance is \mbox{2 
$\times$ 10$^{-5}$} with an error of a factor of $\sim$12. This calculation is based on the observation of one line (145.525 $\mu$m), 
from the "Water In Star forming regions with \textit{Herschel}" (WISH) Project, made with PACS (Woojin Kwon, private 
communication).

%
\subsubsection{Models with a jump at 230\,K with a higher inner abundance} \label{jump230}
%
%

In this section we discuss two species, HCN and HNC (fourth part of Table~\ref{table:Ratran-abundances}). 
{\bbf The collisional rate coefficients for HCN and HNC with H$_2$ were adopted from the \textsc{lamda} database, where they 
were scaled from values with He by \citet{Dumouchel2010}.}

To estimate the abundances of HNC, HCN, and their isotopologues, a model with an abundance jump at 230\,K was adopted. 
\citet{Boonman2001} used a model with a jump at 230\,K, because this temperature was predicted by chemical models as the 
temperature above which most of the atomic oxygen is driven into water. As a result abundances of C- and N-bearing species are 
higher, and derived abundances are \mbox{5 $\times$ 10$^{-6}$} and \mbox{5 $\times$ 10$^{-7}$ for HCN,} and \mbox{1 $\times$ 
10$^{-7}$} and \mbox{1 $\times$ 10$^{-8}$} for HNC, in the inner \mbox{(where $T < 230$\,K)} and outer regions of the envelope. 
The abundance uncertainty is estimated to be within a factor of~4 for HCN and 5 for HNC.
%
%
These results are similar to \citet{Boonman2001} who found 
\mbox{[HCN] = $1 \times 10^{-6}$} in the inner, hot envelope and \mbox{[HCN] = $1 \times 
10^{-8}$} in the outer part of the envelope.
We conclude that jump models reproduce observational lines of 
HCN much better than models with a constant abundance, especially when multiple transitions are observed.

%
\subsubsection{Models with a jump at 100\,K with a higher inner abundance} \label{jump100}
%
%
In this section we discuss species (NH$_3$, SO, SO$_2$, H$_2$S, H$_2$O, HCl, and CH$_3$OH) modeled with an abundance jump at 
100\,K, with a higher inner abundance (the last part of Table~\ref{table:Ratran-abundances}).

The collisional rate coefficients for NH$_3$ {\bbf with H$_2$} were adopted from \citet{Danby1988}. More accurate potentials 
exist \citep{Maret2009}, but they cover a smaller temperature range and do not include hyperfine structure. For ammonia, the 
best-fit model was with an abundance jump at $\sim$100\,K. Models with the same abundances for o-NH$_3$ and p-NH$_3$ were run 
separately.  The expected ortho-to-para ratio is 1:1 for NH$_3$, so both abundances were added to get a total ammonia abundance. From 
the \textsc{ratran} calculations we derived, with an uncertainty of a factor of 4, an ammonia abundance of \mbox{[NH$_3$] = 4 
$\times$ 10$^{-6}$} in the inner part of the envelope \mbox{(where  $T >$ 100\,K)} and \mbox{[NH$_3$] = 4 $\times$ 10$^{-10}$} in 
the outer region of the envelope \mbox{(where  $T <$ 100\,K).} Previously, \citet{Takano1986} derived \mbox{[NH$_3$] $\sim$ 10$^{-8}$} 
by assuming a constant abundance. A model with a jump seems to be more accurate for our data set.

The rate coefficients for collisions of SO with H$_2$ were scaled from collisions of SO with He, which were taken from 
\citet{Lique2006A&A450399L}. Based on 22 SO transitions from a broad range of energy levels \mbox{(\Eup\,$= 26 - 405$\,K)} and 
three $^{34}$SO lines we derived an abundance of \mbox{5 $\times$ 10$^{-7}$} in the inner part of the envelope and \mbox{5 
$\times$ 10$^{-9}$} in the outer region of the envelope \mbox{(where $T < 100$\,K).} The uncertainty of this estimation is within a 
factor of~4. \citet{vanderTak2003}, using six lines from energy levels \mbox{\Eup\,$= 35 - 88$\,K,} estimated a constant SO 
abundance of \mbox{1 $\times$ 10$^{-8}$.} A model with an abundance jump is necessary to fit our SO lines from a broad energy 
range.

The collisional rate coefficients for SO$_2$ {\bbf with H$_2$} were taken from \citet{Green1995}. The most recent {\new calculations} 
of collisional rates were made by \citet{Cernicharo2011}, but they included only temperatures up to 30\,K. This calculation is 
limited to cold interstellar clouds; however, it shows that collisions with molecular hydrogen are about 3 times more effective in 
exciting SO$_2$ than helium, even after scaling for the reduced mass, because the long-range interaction is significant. Thus, we 
scaled the data from \citet{Green1995} by multiplying them by a factor of~3. Based on 50 observed SO$_2$ lines from a broad range of 
energy levels \mbox{(\Eup\,$= 31 - 354$\,K)} and five $^{34}$SO$_2$ transitions we derived an abundance of \mbox{5 $\times$ 
10$^{-6}$} in the inner part of the envelope and \mbox{5 $\times$ 10$^{-9}$} in the outer region of the envelope. The uncertainty 
of this estimation is within a factor of~5. \citet{vanderTak2003} estimated a constant abundance of \mbox{[SO$_2$] = 2 $\times$ 
10$^{-9}$} that is similar to our abundance from the outer part of the envelope; however, the jump is needed to fit our 
observational data.

We scaled the collisional rate coefficients for H$_2$S from the rates for H$_2$O, which were taken from 
\citet{Dubernet2006A&A460323D} and \citet{Daniel2011A&A536A76D}. The ortho-to-para ratio is 3:1 for H$_2$S; thus, models for 
o-H$_2$S and p-H$_2$S were run separately and then both abundances were added to get a total H$_2$S abundance. We derived an 
abundance of \mbox{8 $\times$ 10$^{-8}$} in the inner part of the envelope and \mbox{8 $\times$ 10$^{-10}$} in the outer region of 
the envelope. The uncertainty of this estimation is within a factor of~5. \citet{vanderTak2003}, based on three lines from 
\mbox{\Eup\,$= 28 - 84$\,K,} estimated \mbox{[H$_2$S] = 8 $\times$ 10$^{-9}$.} A model with a jump seems to be more accurate for 
our data set.

The collisional rate coefficients for H$_2$O were taken from \cite{Dubernet2006A&A460323D} and \cite{Daniel2011A&A536A76D}. To get 
a total water abundance, we added results from two models for o-H$_2$O and p-H$_2$O, assuming an o/p ratio of 3:1. The estimated H$_2$O 
abundance is \mbox{1 $\times$ 10$^{-4}$} in the inner part of the envelope and \mbox{1 $\times$ 10$^{-9}$} in the outer region of 
the envelope. The uncertainty of this estimation is within a factor of~4. \citet{vanderTak2006}, from a constant abundance model, 
derived \mbox{[H$_2$O] = 6 $\times$ 10$^{-5}$} and from a jump model the value of \mbox{1.4 $-$ 2 $\times$ 10$^{-4}$} and 
\mbox{10$^{-6}$ $-$ 10$^{-8}$} in the inner and outer part of the envelope, respectively. Both results are very similar, although 
they are based on different physical models. See \citet{Choi2591} for further discussion of the H$_2$O abundance
in the envelope of AFGL 2591.

{\bbf The collisional rate coefficients for HCl with H$_2$ were adopted from the \textsc{lamda} database, where they were 
scaled from the rates for HCl with He by \citet{Neufeld1994ApJ432158N}.} We derived an abundance of \mbox{4 $\times$ 
10$^{-6}$} in the inner part of the envelope and \mbox{4 $\times$ 10$^{-10}$} in the outer region of the envelope. This estimation 
is based on only five detected HCl lines, three hyper-fine components from the energy level of \mbox{\Eup $=$ 30\,K} and two from 
the higher state \mbox{\Eup\, $=$\, 90.1\,K,} thus the uncertainty of this estimation is higher, within a factor of~8.

The collisional rates for  CH$_3$OH {\bbf with H$_2$} were taken from \citet{Rabli2010MNRAS40695R}. Data are 
available for both \mbox{A- and E-type CH$_3$OH,} but given the many CH$_3$OH lines in our spectra, we used only transitions of A-type CH$_3$OH, 
assuming a ratio of \mbox{A$/$E $=$ 1:1}, which is appropriate at the high ($>$30\,K) temperature in the AFGL~2591 envelope. To 
obtain the total methanol abundance we multiplied  our estimate for A-type CH$_3$OH by~2. We estimated an abundance of \mbox{8 
$\times$ 10$^{-7}$} and \mbox{8 $\times$ 10$^{-8}$} in the inner and outer part of the envelope, respectively. The uncertainty of 
this estimation is within a factor of~4. \citet{vanderTak2000a} derived a slightly lower CH$_3$OH abundance, \mbox{8 $\times$ 
10$^{-8}$} in the inner part of the envelope and \mbox{2 $\times$ 10$^{-9}$} in the outer part of the envelope. This calculation 
is based on old collisional rate coefficients and an old physical model.

%
\section{Theoretical modeling} \label{chem}
%
%
%
\subsection{Chemical model} \label{sec:chem_model}
%
%
Our chemical model is based on the advanced time-dependent chemical kinetics \textsc{alchemic} code \citep[see][]{Semenov_ea10}. 
The gas-grain deuterium chemistry network of \citet{Albertsson_ea13,Albertsson_ea14a} is utilized. The non-deuterated chemical 
network is based on the osu.2007 rate file \footnote{\tt http://faculty.virginia.edu/ericherb/research.html}.
The network is supplied with a set of approximately $1\,000$ reactions with high-temperature barriers from \citet{Harada_ea10} and 
\citet{Harada_ea12}. The recent updates as of June 2013 to the reaction rates are implemented (e.g., from the \textsc{kida} 
database\footnote{\tt http://kida.obs.u-bordeaux1.fr}; see \citet{KIDA} and \citet{Albertsson_ea13}).

Primal isotope exchange reactions for H$_3^+$ as well as CH$_3^+$ and C$_{2}$H$_{2}^+$ from \citet{2000A&A...361..388R}, 
\citet{GHR_02}, \citet{2004A&A...424..905R}, and \citet{2005A&A...438..585R} were included. In cases where the position of the 
deuterium atom in a reactant or in a product was ambiguous, a statistical branching approach was used. In \citet{Albertsson_ea14a} 
this deuterium network was further extended by adding ortho and para forms of H$_2$, H$_2^+$, and H$_3^+$ isotopologues and the 
related nuclear spin-state exchange processes from several experimental and theoretical studies 
\citep{1990JChPh..92.2377G,GHR_02,2004A&A...427..887F,2004A&A...418.1035W,2006A&A...449..621F,2009A&A...494..623P,
2009JChPh.130p4302H,2011PhRvL.107b3201H,2013A&A...554A..92S}.
 
In our standard model, we used the cosmic ray (CR) ionization rate \mbox{$\zeta_{\rm CR}=5 \times 10^{-17}$~s$^{-1}$,} as derived 
by \citet{Tak2000c} specifically for AFGL 2591.
In addition, we considered two other models where the CR ionization rate was increased by factors of 10 and~100.

Several tens of newer photoreaction rates are adopted from  \citet{vDea_06}\footnote{\tt 
http://www.strw.leidenuniv.nl/$\sim$ewine/photo}. In the standard model, the AFGL~2591 envelope is assumed to be bathing in an 
ambient interstellar FUV field with an intensity of one \citet{G} interstellar UV. In addition, we considered two other models 
where the UV intensity (in Draine's unit of $\chi_0$) was assumed to be 50 and 500. The self- and mutual-shielding of CO and H$_2$ from 
external dissociating radiation was calculated as in \citet{Semenov_Wiebe11}. 

The grains are uniform amorphous olivine particles with a density of $3$~g\,cm$^{-3}$ and a radius of $0.1\,\mu$m. Each grain 
provides around $1.88\times10^6$ surface sites \citep[][]{Bihamea01}. Two models with smaller ($0.03\,\mu$m) and bigger 
($1\,\mu$m) uniform grains were also considered.

The gas-grain interactions include sticking of neutral species and electrons to dust grains with 100\% probability and desorption 
of ices by thermal, CR-, and UV-driven processes. 
We do not allow H$_2$ isotopologues to stick to grains because for H$_2$ this requires temperatures of $\la 4$~K. A UV photodesorption 
yield of $3\times 10^{-3}$ was adopted \citep[e.g.,][]{2009A&A...504..891O,2009A&A...496..281O,Fayolle_ea11a,2013A&A...556A.122F}. 
Photodissociation processes of solid species are taken from \citet{Garrod_Herbst06} and \citet{Semenov_Wiebe11}. 

Surface recombination proceeds through the classical Langmuir-Hinshelwood mechanism \citep[e.g.,][]{HHL92}. The ratio between 
diffusion and desorption energies of surface reactants is taken to be 0.7 for all surface species {\new \citep{ruffle2000}}. We do not allow tunneling of 
surface species via the potential wells of the adjacent surface sites. To account for hydrogen tunneling through barriers of 
surface reactions, we  employed Eq.~(6) from \citep{HHL92}, which describes a tunneling probability through a rectangular 
barrier with thickness of $1\, \AA$.

For each surface recombination, we assume a $1\,\%$ probability for the products to leave the grain as a result of  the partial 
conversion of the reaction exothermicity into breaking the surface-adsorbate bond \citep{2007A&A...467.1103G,2013ApJ...769...34V}. 
Following experimental studies on the formation of molecular hydrogen on amorphous dust grains by \citet{Katz_ea99}, the standard 
rate equation approach to the surface chemistry is adopted. In addition, dissociative recombination and radiative neutralization 
of molecular ions on charged grains and grain recharging are taken into account.

Overall, the chemical network consists of 1268 species made of 13 elements and 38812 reactions. With this network and using 
$10^{-5}$ relative errors and $10^{-15}$ absolute errors, the AFGL~2591 model takes about 3~minutes of CPU time (Core-i7 2.0~GHz, 
gfortran~4.8-x64) to calculate over an evolutionary time span of \mbox{$1-10$ $\times\,10^4$}~years (with 99 logarithmic time 
steps), which was proposed by \citet{Stauber2005} as the chemical age of AFGL~2591. 

%
\subsection{Initial abundances} \label{sec:in_abunds}
%
%

\begin{table}
\setlength{\extrarowheight}{2pt}
\caption{Initial abundances {\new (relative to total H)} for modeling pre-AFGL~2591 evolutionary phase.}             
\label{tab:init_abunds_pre}
\centering                          
\begin{tabular}{ll|ll}        
\hline\hline
Species & Abundances &Species & Abundances\\               
\hline                        
\multicolumn{4}{c}{Low metals  (\textsc{lm}), ${\rm C/O} = 0.44$}\\
\hline
o-H$_2$ &   $3.75\, \times$ 10$^{-1}$&S       &   $9.14\, \times$ 10$^{-8}$  \\
p-H$_2$ &   $1.25\, \times$ 10$^{-1}$&Si      &   $9.74\, \times$ 10$^{-9}$ \\
HD      &   $1.55\, \times$ 10$^{-5}$&Na      &   $2.25\, \times$ 10$^{-9}$ \\
He      &   $9.75\, \times$ 10$^{-2}$&Mg      &   $1.09\, \times$ 10$^{-8}$\\
C       &   $7.86\, \times$ 10$^{-5}$&Fe      &   $2.74\, \times$ 10$^{-9}$ \\
N       &   $2.47\, \times$ 10$^{-5}$&P       &   $2.16\, \times$ 10$^{-10}$ \\
O       &   $1.80\, \times$ 10$^{-4}$&Cl      &   $1.00\, \times$ 10$^{-9}$  \\
\hline                        
\multicolumn{4}{c}{Dutrey et al.~(2011), their model C (\textsc{d11c}), ${\rm C/O} = 1.21$}\\
\hline
o-H$_2$ &   $3.75\, \times$ 10$^{-1}$  &Si$^+$      &   $8.00\, \times$ 10$^{-9}$\\
p-H$_2$ &   $1.25\, \times$ 10$^{-1}$  &Na$^+$      &   $2.00\, \times$ 10$^{-9}$ \\
HD      &   $1.55\, \times$ 10$^{-5}$  &Mg$^+$      &   $7.00\, \times$ 10$^{-9}$\\
He      &   $9.00\, \times$ 10$^{-2}$  &Fe$^+$      &   $3.00\, \times$ 10$^{-9}$\\
C$^+$       &   $1.7\, \times$ 10$^{-4}$  &P$^+$      &   $2.00\, \times$ 10$^{-10}$ \\
N       &   $6.2\, \times$ 10$^{-5}$  &Cl$^+$      &   $1.00\, \times$ 10$^{-9}$ \\
O       &   $1.40\, \times$ 10$^{-4}$  &e$^-$      &   $1.77\, \times$ 10$^{-4}$\\
S$^+$       &   $8.00\, \times$ 10$^{-9}$   \\
\hline                        
\multicolumn{4}{c}{Hincelin et al.~(2011) model (\textsc{h11}), ${\rm C/O} = 1.13$}\\
\hline
o-H$_2$ &   $3.75\, \times$ 10$^{-1}$  &Si$^+$      &   $8.00\, \times$ 10$^{-9}$ \\
p-H$_2$ &   $1.25\, \times$ 10$^{-1}$  &Na$^+$      &   $2.00\, \times$ 10$^{-9}$\\
HD      &   $1.55\, \times$ 10$^{-5}$  &Mg$^+$      &   $7.00\, \times$ 10$^{-9}$ \\
He      &   $9.00\, \times$ 10$^{-2}$  &Fe$^+$      &   $3.00\, \times$ 10$^{-9}$\\
C$^+$       &   $1.7\, \times$ 10$^{-4}$  &P$^+$      &   $2.00\, \times$ 10$^{-10}$\\
N       &   $6.2\, \times$ 10$^{-5}$  &Cl$^+$      &   $1.00\, \times$ 10$^{-9}$\\
O       &   $1.50\, \times$ 10$^{-4}$  &F$^+$       &   $6.68\, \times$ 10$^{-9}$ \\
S$^+$       &   $8.00\, \times$ 10$^{-8}$   &e$^-$      &   $1.70\, \times$ 10$^{-4}$\\
\hline                                   
\end{tabular}
\end{table}

\begin{table}
\setlength{\extrarowheight}{2.3pt}
\caption{The top 25 of the initially abundant molecules {\new (relative to total H)} for the AFGL~2591 chemical modeling.}
\label{tab:init_abunds_afgl}
\centering                          
\begin{tabular}{ll|ll}        
\hline\hline
Species & Abundances & Species & Abundances\\
\hline                        
\multicolumn{4}{c}{\textsc{lm} model with ${\rm C/O} = 0.44$}\\
\hline                        
p-H$_2$     &   $3.77\, \times$ 10$^{-1}$ &CH$_4^*$    &   $3.64\, \times$ 10$^{-6}$\\
o-H$_2$     &   $1.23\, \times$ 10$^{-1}$ &N$_2^*$     &   $1.76\, \times$ 10$^{-6}$\\
He          &   $9.75\, \times$ 10$^{-2}$ &H$^*$       &   $6.03\, \times$ 10$^{-7}$\\
H           &   $5.25\, \times$ 10$^{-4}$ &C$_3$H$_2^*$&   $4.48\, \times$ 10$^{-7}$  \\
H$_2$O$^*$  &   $5.53\, \times$ 10$^{-5}$ &OH          &   $3.43\, \times$ 10$^{-7}$\\
CO$^*$      &   $4.05\, \times$ 10$^{-5}$ &H$_2$O      &   $2.79\, \times$ 10$^{-7}$\\
CO          &   $3.26\, \times$ 10$^{-5}$ &HNO$^*$     &   $2.40\, \times$ 10$^{-7}$ \\
O$_2$       &   $1.79\, \times$ 10$^{-5}$ &NO          &   $2.22\, \times$ 10$^{-7}$ \\
HD          &   $1.52\, \times$ 10$^{-5}$ &N           &   $1.36\, \times$ 10$^{-7}$  \\
N$_2$       &   $7.39\, \times$ 10$^{-6}$ &HDO$^*$     &   $1.35\, \times$ 10$^{-7}$  \\
NH$_3^*$    &   $5.64\, \times$ 10$^{-6}$ &CO$_2$      &   $1.32\, \times$ 10$^{-7}$  \\
O           &   $5.59\, \times$ 10$^{-6}$ &CO$_2^*$    &   $1.19\, \times$ 10$^{-7}$ \\
O$_2^*$     &   $4.12\, \times$ 10$^{-6}$  \\
\hline                        
\multicolumn{4}{c}{\textsc{d11c} model with ${\rm C/O} = 1.21$}\\
\hline                        
o-H$_2$     &   $3.06\, \times$ 10$^{-1}$&C$_2$H$_6^*$&   $6.92\, \times$ 10$^{-6}$\\\
p-H$_2$     &   $1.93\, \times$ 10$^{-1}$&N$_2$       &   $6.07\, \times$ 10$^{-6}$\\
He          &   $9.00\, \times$ 10$^{-2}$&H$_2$CO$^*$ &   $5.76\, \times$ 10$^{-6}$ \\
CO$_2^*$    &   $3.11\, \times$ 10$^{-5}$&HCN$^*$     &   $4.87\, \times$ 10$^{-6}$ \\
CH$_4^*$    &   $3.05\, \times$ 10$^{-5}$&CO$^*$      &   $4.45\, \times$ 10$^{-6}$\\
H           &   $2.57\, \times$ 10$^{-5}$&HNC$^*$     &   $1.79\, \times$ 10$^{-6}$ \\
CH$_3$OH$^*$&   $2.57\, \times$ 10$^{-5}$&CH$_5$N$^*$ &   $1.28\, \times$ 10$^{-6}$ \\
NH$_3$      &   $2.39\, \times$ 10$^{-5}$&C$_3$       &   $9.27\, \times$ 10$^{-7}$ \\
H$_2$O$^*$  &   $2.30\, \times$ 10$^{-5}$&HNCO$^*$    &   $5.64\, \times$ 10$^{-7}$  \\
CO          &   $1.69\, \times$ 10$^{-5}$&CH$_3$D$^*$ &   $4.11\, \times$ 10$^{-7}$ \\
HD          &   $1.42\, \times$ 10$^{-5}$&CH$_4$      &   $2.82\, \times$ 10$^{-7}$ \\
N$_2^*$     &   $7.91\, \times$ 10$^{-6}$&N           &   $2.82\, \times$ 10$^{-7}$ \\
C$_3$H$_4^*$&   $7.89\, \times$ 10$^{-6}$ \\
\hline                        
\multicolumn{4}{c}{\textsc{h11} model with ${\rm C/O} = 1.13$}\\
\hline                        
p-H$_2$       &   $3.97\, \times$ 10$^{-1}$&C$_2$H$_6^*$  &   $6.90\, \times$ 10$^{-6}$ \\
o-H$_2$       &   $1.01\, \times$ 10$^{-1}$&H$_2$CO$^*$   &   $6.02\, \times$ 10$^{-6}$ \\
He            &   $9.00\, \times$ 10$^{-2}$&N$_2$         &   $5.95\, \times$ 10$^{-6}$ \\
CO$_2^*$      &   $3.45\, \times$ 10$^{-5}$&HCN$^*$       &   $5.06\, \times$ 10$^{-6}$ \\
CH$_4^*$      &   $2.91\, \times$ 10$^{-5}$&CO$^*$        &   $4.86\, \times$ 10$^{-6}$ \\
H             &   $2.59\, \times$ 10$^{-5}$&HNC$^*$       &   $1.81\, \times$ 10$^{-6}$ \\
H$_2$O        &   $2.57\, \times$ 10$^{-5}$&CH$_5$N$^*$   &   $1.31\, \times$ 10$^{-6}$ \\
CH$_3$OH$^*$  &   $2.44\, \times$ 10$^{-5}$&C$_3$         &   $8.03\, \times$ 10$^{-7}$ \\
NH$_3^*$      &   $2.42\, \times$ 10$^{-5}$&HNCO$^*$      &   $6.47\, \times$ 10$^{-7}$ \\
CO            &   $1.76\, \times$ 10$^{-5}$&CH$_3$D$^*$   &   $3.97\, \times$ 10$^{-7}$ \\
HD            &   $1.37\, \times$ 10$^{-5}$&N             &   $3.63\, \times$ 10$^{-7}$  \\
N$_2^*$       &   $7.71\, \times$ 10$^{-6}$&O             &   $3.57\, \times$ 10$^{-7}$  \\
C$_3$H$_4^*$  &   $7.40\, \times$ 10$^{-6}$  \\
\hline                        
\end{tabular}
  \begin{list}{}{}
        \item[* -- ] denotes frozen species.
   \end{list} 
\end{table} 

To set the initial abundances of the model, we calculated the chemical evolution of a 0D molecular cloud 
with $n_{\rm H} = 2\times10^4$~cm$^{-3}$, $T  = 10$~K, and $A_{\rm V}=10$~mag over 1~Myr. The neutral "low metals" 
(model \textsc{lm}) elemental abundances of \citet{1982ApJS...48..321G}, \citet{Lea98}, and \citet{2013ChRv1138710A} were 
used, with the solar ratio of ${\rm C/O} = 0.44$, initial ortho/para H$_2$ of 3:1, hydrogen being fully in molecular form, and 
deuterium locked in HD (see Table~\ref{tab:init_abunds_pre}). The resulting abundances of modeling this pre-AFGL~2591 
phase were used as initial abundances for the AFGL~2591 chemical modeling (see Table~\ref{tab:init_abunds_afgl}).
{\new The instantaneous jump from a 0D to a 1D structure is a crude approximation which future work should address in more detail.}

%
%
%
In addition {\new to the \textsc{lm} abundance set}, we  considered several initial molecular abundance sets with non-solar C/O ratios, which were proposed to 
explain observed peculiarities of the sulfur and oxygen chemistries in high-mass star-forming regions and protoplanetary disks 
\citep{2011A&A...530A..61H,2011A&A...535A.104D}, called \textsc{d11c} and \textsc{h11} (see Table~\ref{tab:init_abunds_pre}). Both 
these elemental abundance sets have ${\rm C/O} > 1$, an  abundance of nitrogen 3 times higher than in our standard \textsc{lm} set, and 
most of the atoms in the ionized state. In addition, the set of Dutrey et al.~(2011) has a 10 times lower sulfur abundance than the \textsc{lm} 
set. 

Similar to our standard \textsc{lm}~model, 
we  calculated the chemical evolution of a 0D molecular cloud to set the initial abundances for the 1D physical-chemical model of 
the AFGL~2591 envelope (see Table~\ref{tab:init_abunds_afgl}) with the two non-solar C/O initial abundances. 
The physical parameters of both models were the same, $n_{\rm H} = 2\times10^5$~cm$^{-3}$, $ T = 
15$~K, and $A_{\rm V}=10$~mag, and we adopted an age of 50~kyr. {\bbf These physical conditions differ from the \textsc{lm} case 
presented above. Since AFGL 2591 is forming a cluster of low- and intermediate-mass stars, it
is not obvious which physical conditions more adequately describe the pre-AFGL 2591
phase, namely, those characteristic of low-mass prestellar cores or those characteristic of
higher-mass infrared dark clouds. The two \textsc{d11c} and \textsc{h11}  models correspond to the high-mass 
case, whereas the \textsc{lm} model corresponds to the low-mass case.}

The initial abundances calculated with this 0D physical model and the 
\citet{2011A&A...535A.104D} (their model C) elemental set is called \textsc{d11c}. Similarly, the model of initial elemental 
abundances of \citet{2011A&A...530A..61H} adopted to calculate \mbox{pre-AFGL~2591} the initial abundances is called \textsc{h11} (see 
Table~\ref{tab:init_abunds_afgl}).

%
\subsection{Error estimations} \label{chem-errors}
%
%
The difficulty of calculating abundance uncertainties is well known in chemical studies of various astrophysical environments, 
ranging from dark clouds and hot cores to protoplanetary disks and exoplanetary atmospheres \citep[see, e.g., 
][]{2013ChRv1138710A,Dobrijevic2003A&A398335D,2004AstL...30..566V,2005A&A...444..883W,Wakelam2006A&A451551W,Vasyunin_ea08}. The 
error budget of the theoretical abundances is determined by both the uncertainties in physical conditions in the object and, to a 
larger degree, by uncertainties in the adopted reaction rate coefficients and their barriers. Poorly known initial conditions for 
chemistry may also play a role here.

In order to estimate the chemical uncertainties rigorously one needs to perform a large-scale Monte Carlo modeling by 
varying the reaction rates within their error bars and recalculating the chemical evolution of the specific astrophysical environment. 
We do not attempt to perform such a detailed study and use the estimates from previous works. 

Previous studies of chemical uncertainties have found that the uncertainties are in general larger for bigger 
molecules because their evolution involves a larger number of reactions compared to simpler molecules. For simpler species such as 
CO and H$_2$ involved in a limited cycle of reactions it is easier to derive the reaction rates with a high accuracy of $\sim 
25\%$. Consequently, their abundances are usually accurate within $10-30\%$ in modern astrochemical models.  On the other hand, 
for other diatomic and triatomic species such as CN, HCO$^+$, HCN, and CCH, the uncertainties are usually about a factor of \mbox{3 
$-$ 4} \citep[see][]{Vasyunin_ea08,2013ChRv1138710A}. The chemical uncertainties are even higher for more complex molecules like 
methanol and probably exceed a factor of 10.

For S-bearing species, for which many reaction rates have not been properly measured or calculated or included in the 
networks, these intrinsic uncertainties and hence the uncertainties in their resulting abundances are higher, a factor of 
$\ga\,10$ even for simple species such as SO, OCS, and SO$_2$ \citep{Loison2012MNRAS}. In addition, the incompleteness of astrochemical 
networks with regard to the chemistry of Cl- and F-bearing molecules makes their calculated abundances less reliable. 

In our study we assume that the uncertainties in the abundances of ortho- and para-H$_2$ and CO are within 30\%. For 
HCO$^+$, H$_2$CO, CN, N$_2$H$^+$, C$_2$H, NO, OH, C, C$^+$, O, CH, NH$_3$, H$_2$O, HCN, and HNC, the uncertainties are about a factor 
of~3, and for S-bearing species, CH$_3$OH, and HCl they are about a factor of 10.

\begin{table}
\setlength{\extrarowheight}{1.8pt}
\caption{Agreement (A) between the observed and best-fit 1D abundance profiles.}
\label{tab:best_fit}
\begin{center}
\begin{tabular}{lc|lc|lc}
\hline\hline
Species & A &Species & A &Species & A \\
\hline
H$_2$     & $+$&OCS       & $+$&NH$_3$    & $+$\\
CO        & $+$&OH        & $+$&SO        & $+$\\
CN        & $+$&H$_2$CS   & $+$&SO$_2$    & $?$\\
CS        & $?$&C        & $?$&H$_2$S    & $?$\\
HCO$^+$   & $+$&C$^+$       & $?$&H$_2$O    & $+$\\
H$_2$CO   & $+$&CH        & $?$&HCl       & $?$\\
N$_2$H$^+$& $+$&O        & $+$&CH$_3$OH  & $-$\\
CCH       & $?$&HCN       & $-$& &\\
NO        & $+$&HNC       & $+$& &\\
\hline
Overall   & \multicolumn{5}{l}{$19/25 = 76\%$}\\
\hline    
\end{tabular}
\end{center}
 \tablefoot{[``$+$''] = good, ``$?$'' = questionable, [``$-$''] -- poor agreement. {\new The best-fit model has {\sc LM} abundances, C/O = 0.44, $\zeta$ = \mbox{5$\times$10$^{-17}$\,s$^{-1}$}, $\chi_0$ = 1, and $t$ = 10--50\,kyr.}}
\end{table}

\begin{table}
\setlength{\extrarowheight}{2pt}
\caption{Sensitivity of assorted molecules to varied parameters of the model, the ionization rate (CR), the C/O ratio, and the 
chemical age.}
\label{tab:species_sens}
\begin{center}
\begin{tabular}{lccc|lccc}
\hline\hline
Species & \multicolumn{3}{c}{Parameters} & Species & \multicolumn{3}{c}{Parameters}\\
        & CR & C/O & Age & & CR & C/O & Age\\
\hline
H$_2$     & $-$     & $-$ & $-$ & C$^+$       & $-$     & $+$ & $-$\\
CO        & $-$     & $+$ & $-$ & CH        & $-$     & $+$ & $?$\\
CN        & $-$     & $+$ & $-$ & O        & $-$     & $+$ & $-$\\
CS        & $-$     & $+$ & $?$ & HCN       & $+$     & $+$ & $+$\\
HCO$^+$   & $?$     & $?$ & $-$ & HNC       & $+$     & $+$ & $+$\\
H$_2$CO   & $+$     & $+$ & $?$ & NH$_3$    & $-$     & $+$ & $+$\\
N$_2$H$^+$& $?$     & $-$ & $-$ & SO        & $-$     & $+$ & $-$\\
CCH       & $-$     & $+$ & $-$ & SO$_2$    & $-$     & $+$ & $-$\\
NO        & $-$     & $?$ & $-$ & H$_2$S    & $-$     & $+$ & $+$\\
OCS       & $-$     & $?$ & $+$ & H$_2$O    & $?$     & $?$ & $-$\\
OH        & $-$     & $?$ & $-$ & HCl       & $-$     & $-$ & $-$\\
H$_2$CS   & $?$     & $+$ & $+$ & CH$_3$OH  & $-$     & $+$ & $?$\\
C        & $-$     & $+$ & $-$ \\
\hline
\end{tabular}
\end{center}
\tablefoot{``$+$'' = strongly sensitive, ``$?$'' = weakly sensitive, ``$-$'' = insensitive to a given parameter.}
\end{table}

\begin{figure*}
\includegraphics[width=0.34\textwidth]{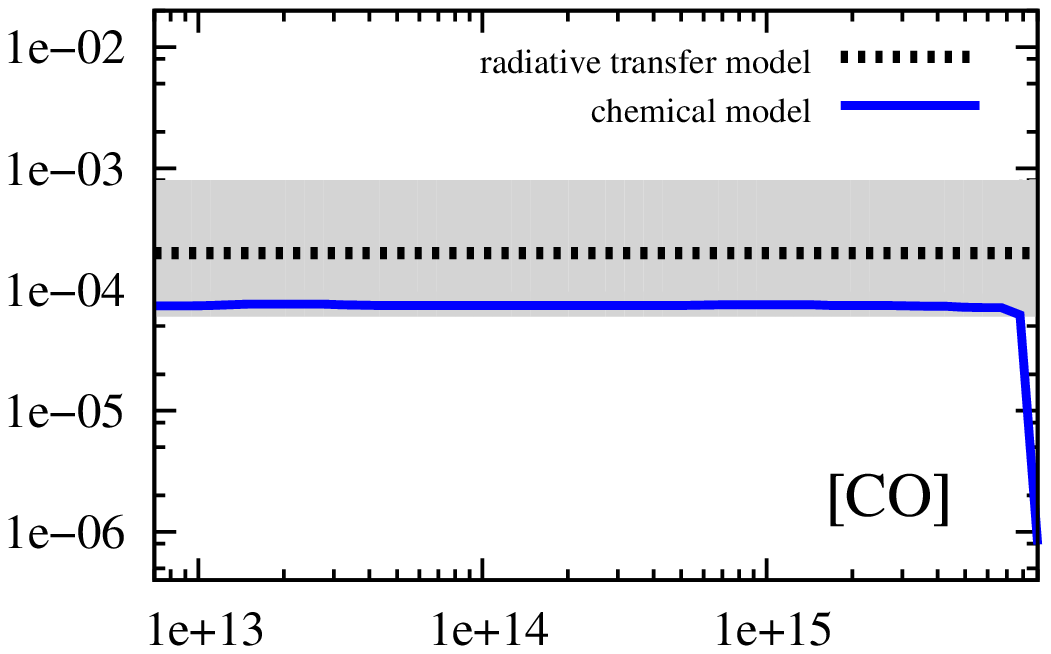}
\includegraphics[width=0.34\textwidth]{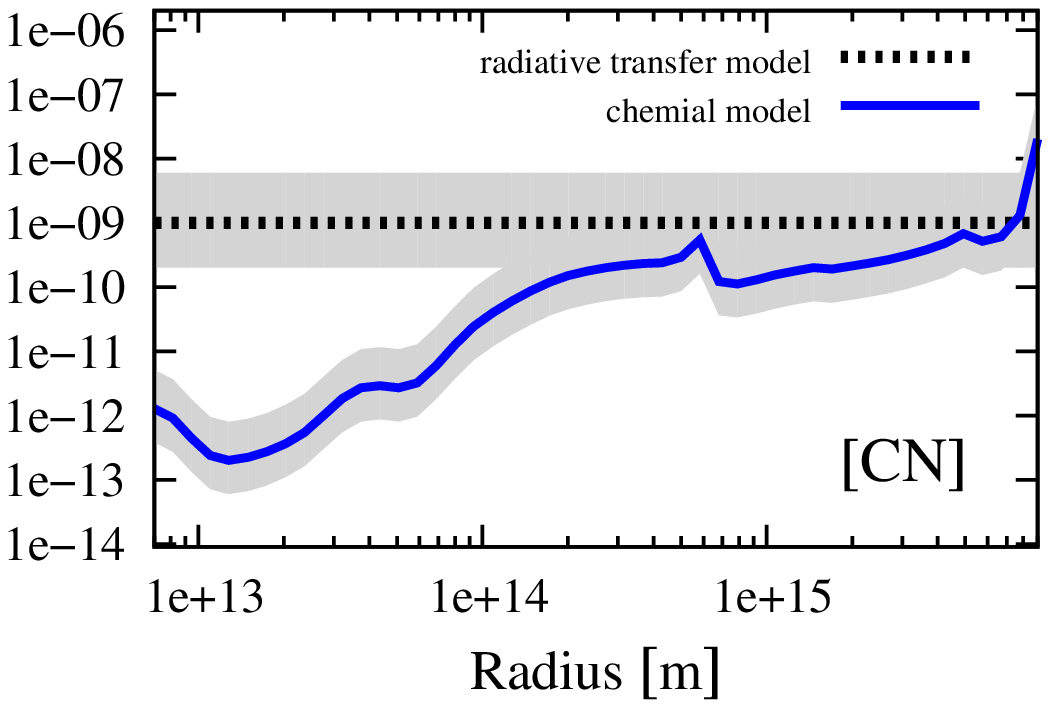}
\includegraphics[width=0.34\textwidth]{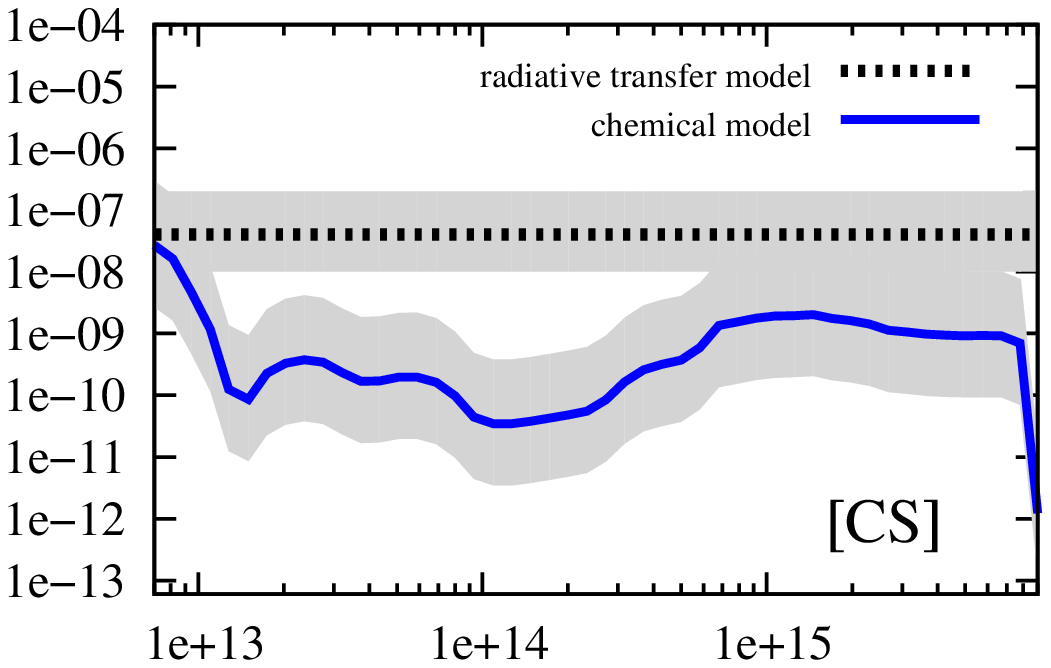}
\caption[]{The radial distribution of abundances with error-bars of CO, CN, and CS at the time of \mbox{$10^4$} years. The 
\textsc{ratran}-derived values (dashed lines) are compared with the results of the best-fit 1D chemical model (solid lines); {\new see Table~\ref{tab:best_fit} for the specifications}.} 
\label{fig:comp1}
\end{figure*}
\begin{figure*}
\includegraphics[width=0.34\textwidth]{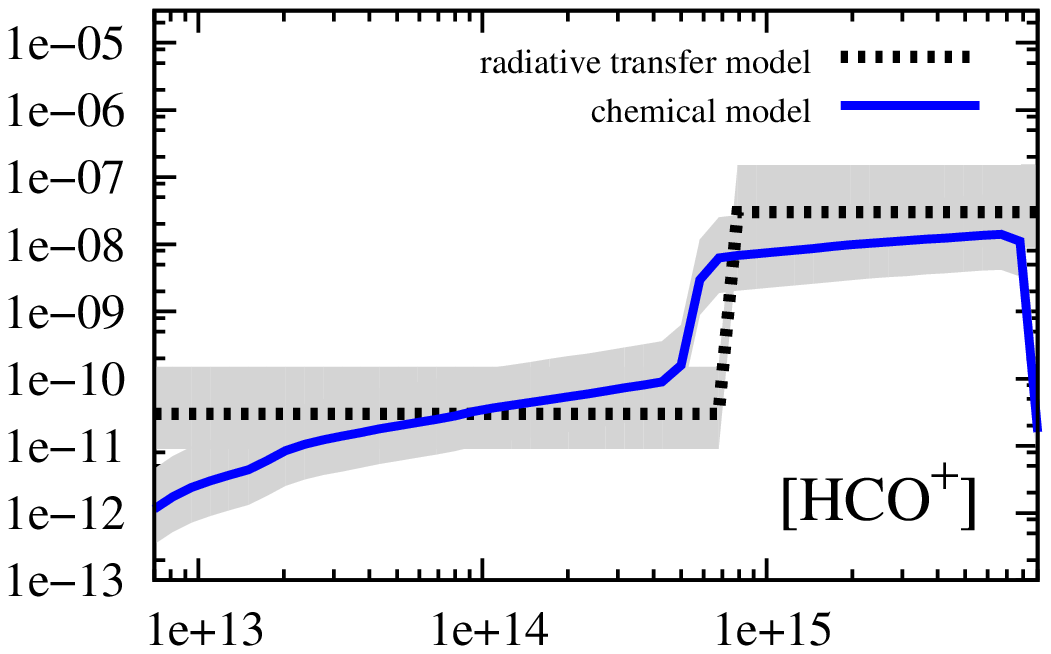}
\includegraphics[width=0.34\textwidth]{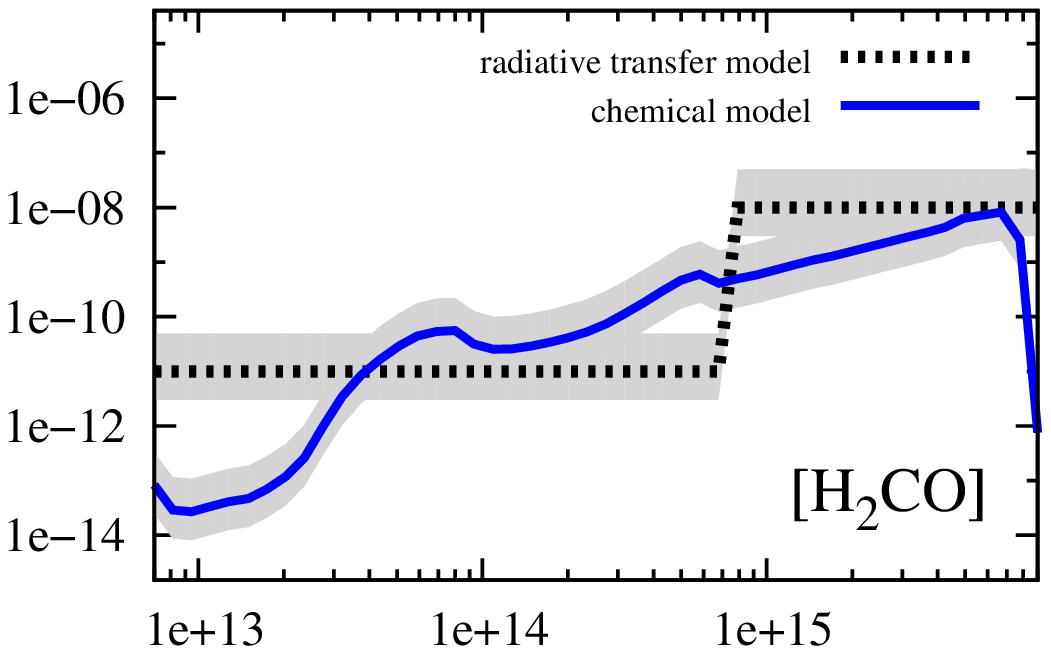}
\includegraphics[width=0.34\textwidth]{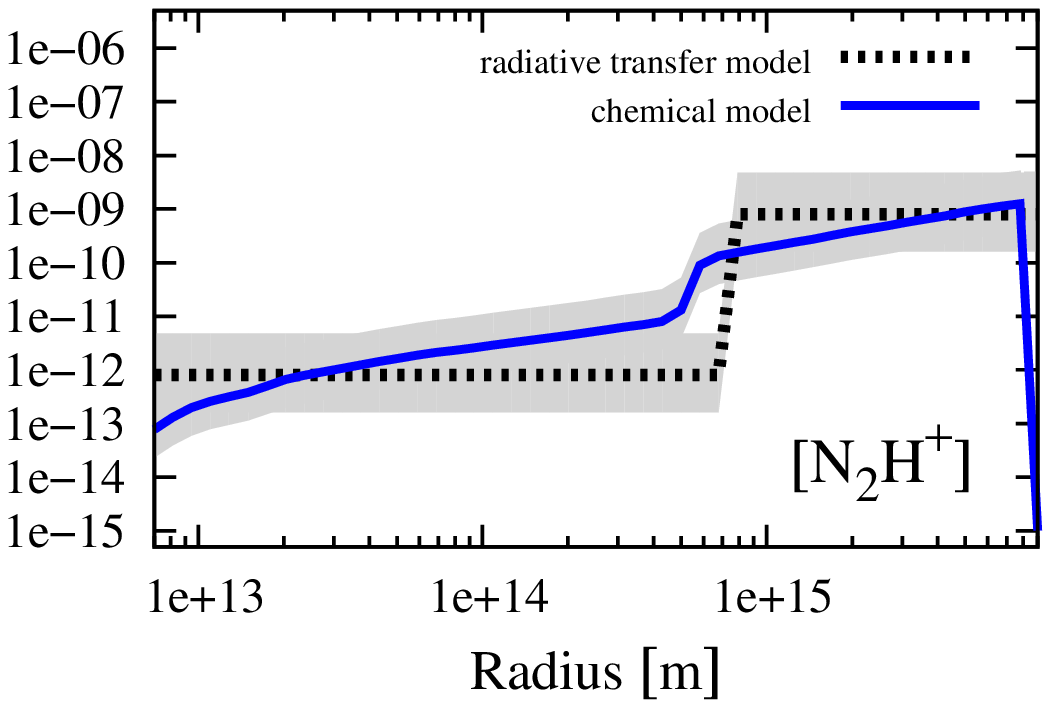}\\
\includegraphics[width=0.34\textwidth]{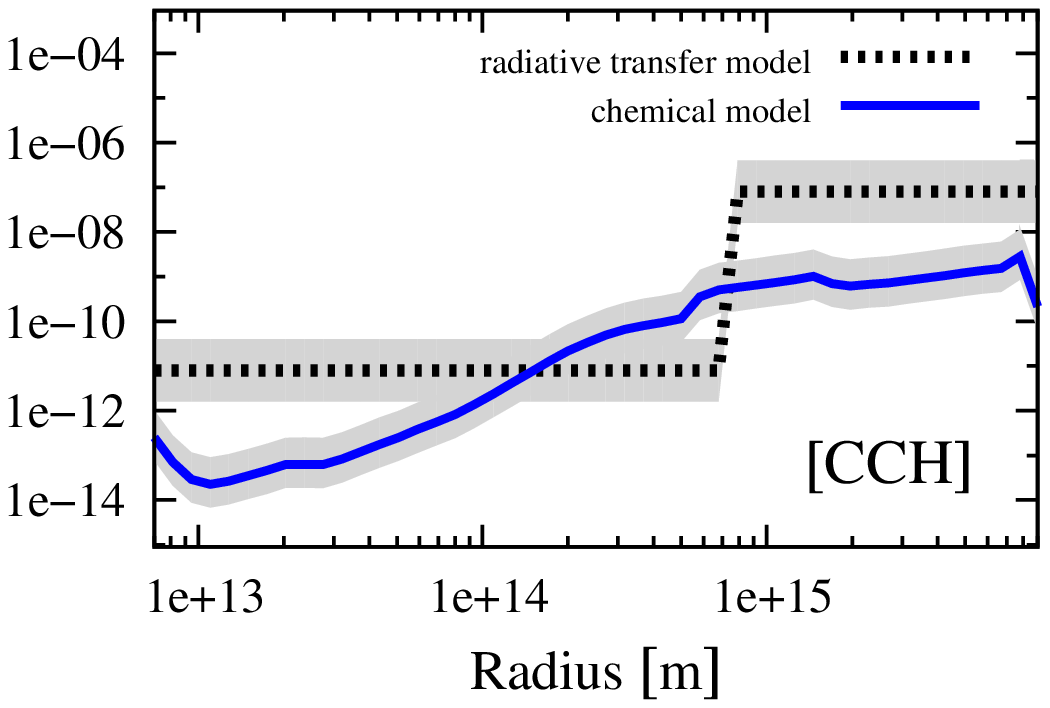}
\caption[]{The same as Fig. 2. Radiative transfer models are with a jump at 100\,K (lower inner abundance)
.} \label{fig:comp1b}
\end{figure*}
\begin{figure*}
\includegraphics[width=0.34\textwidth]{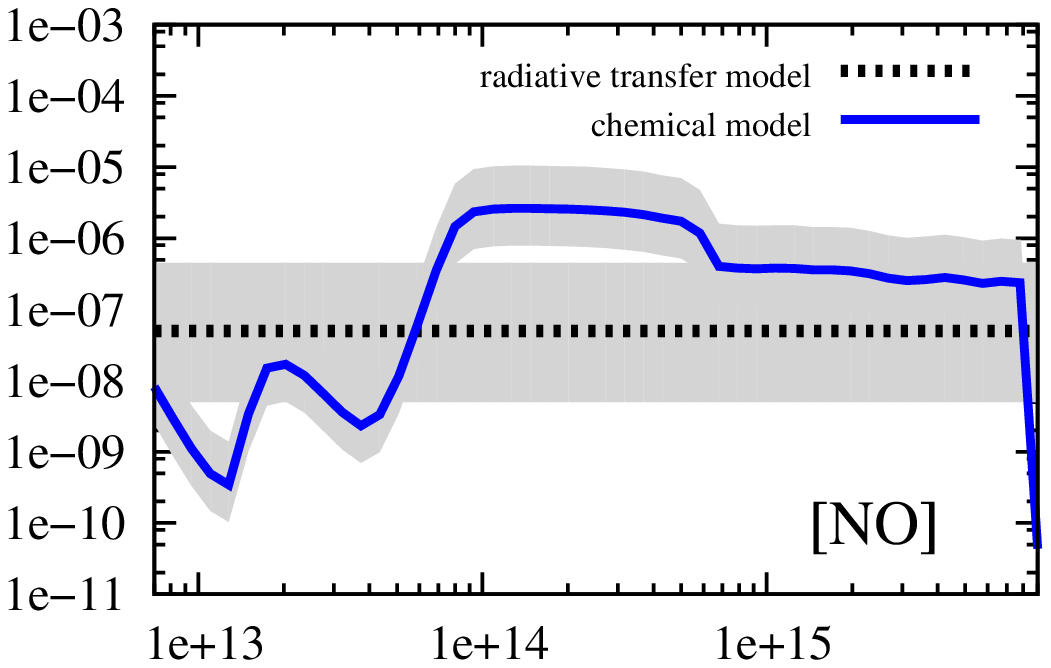}
\includegraphics[width=0.34\textwidth]{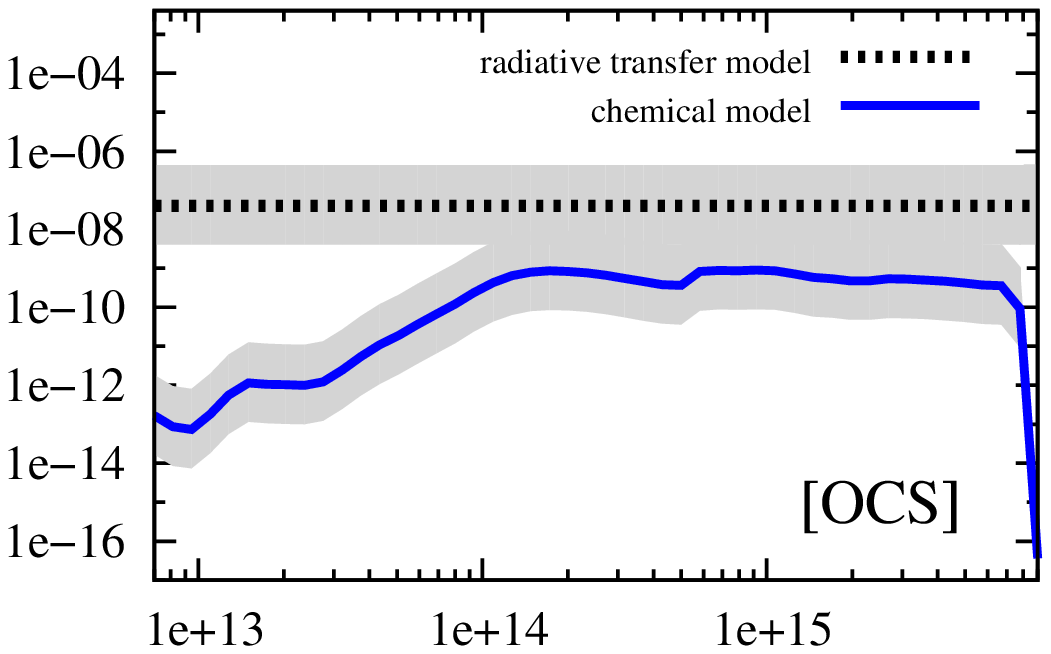}
\includegraphics[width=0.34\textwidth]{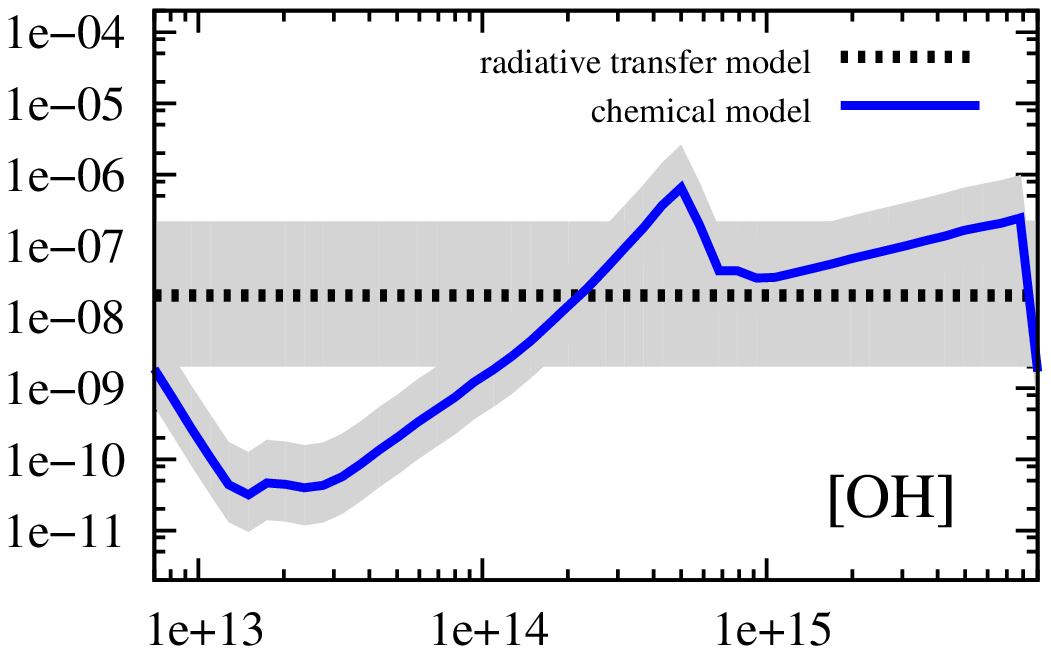}\\
\includegraphics[width=0.34\textwidth]{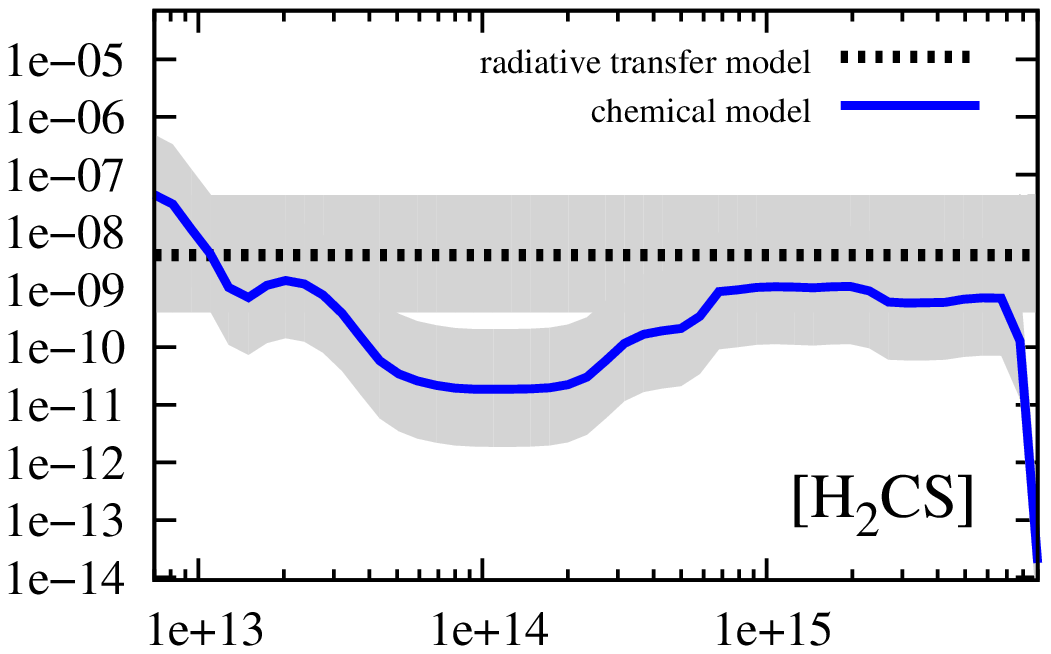}
\includegraphics[width=0.34\textwidth]{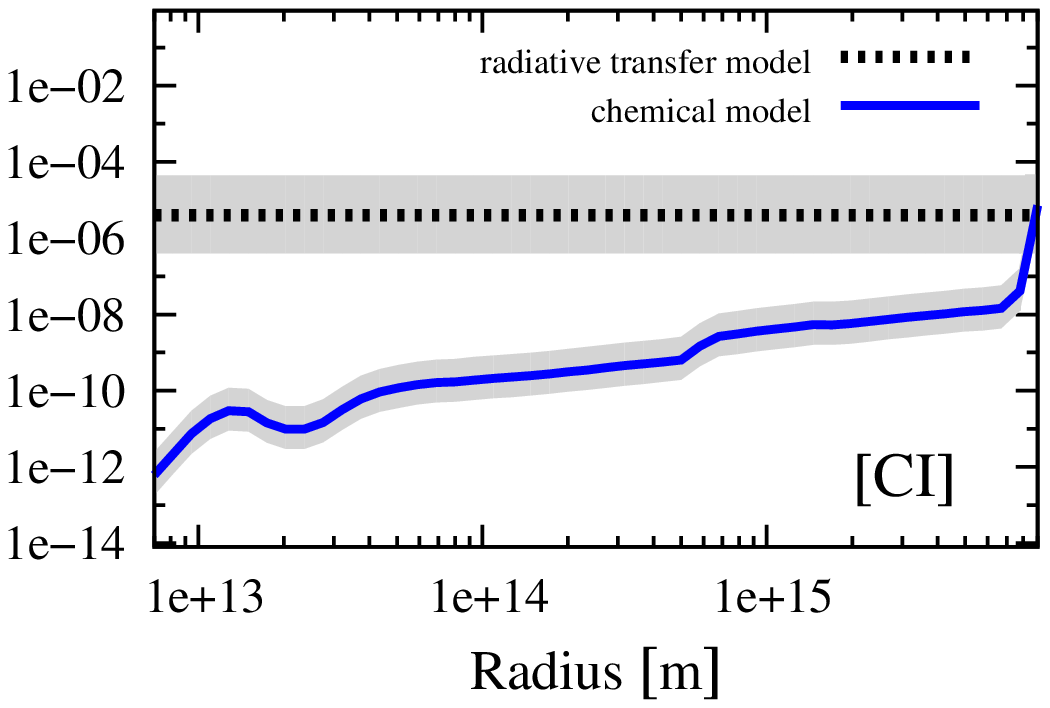}
\includegraphics[width=0.34\textwidth]{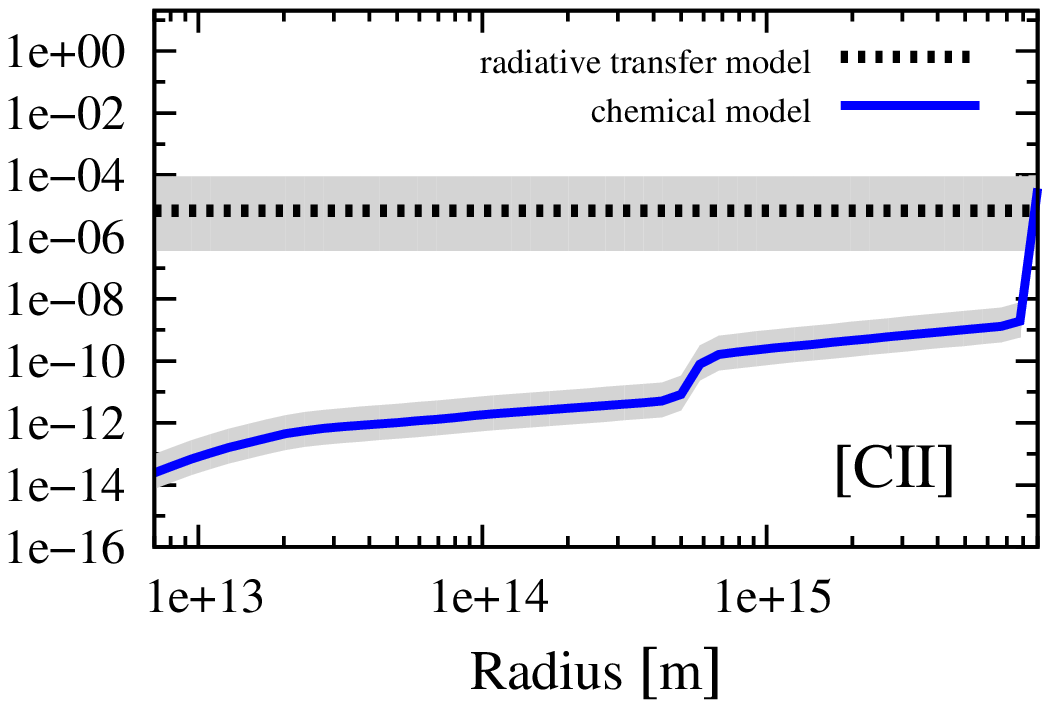}\\
\includegraphics[width=0.34\textwidth]{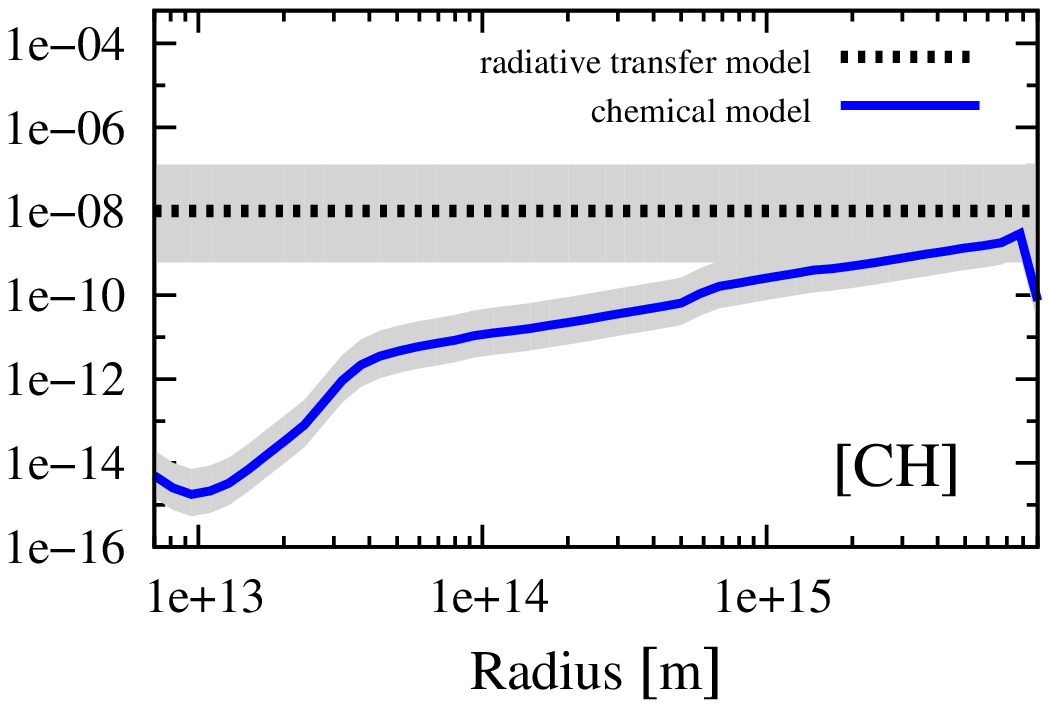}
\includegraphics[width=0.34\textwidth]{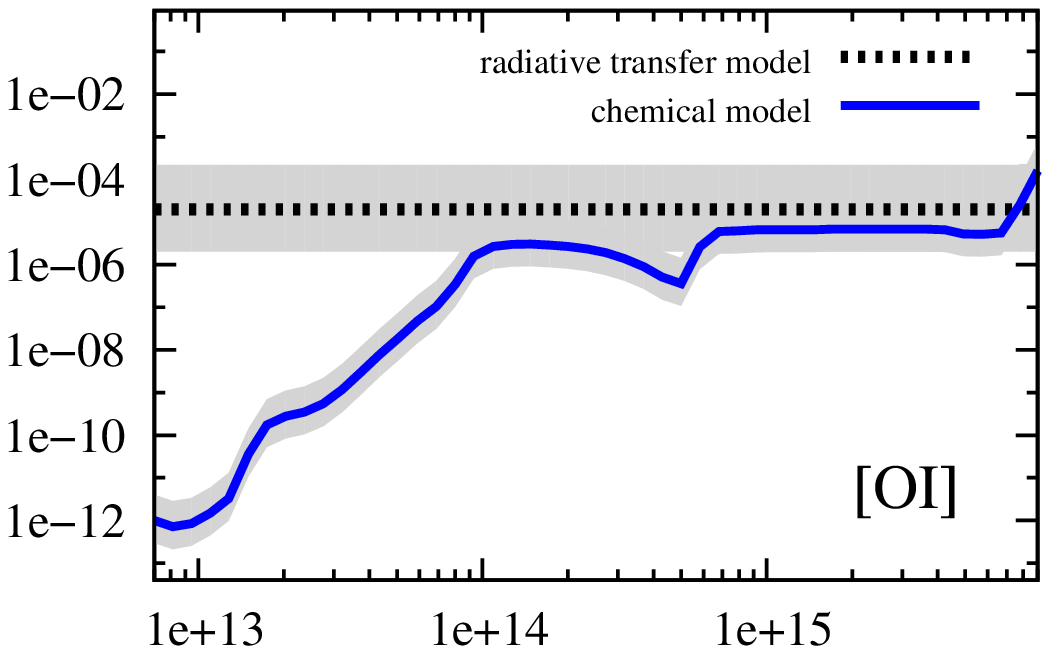}
\caption[]{The same as Fig. 2. Radiative transfer models with constant abundance are based on only a few observed transitions
.} 
\label{fig:comp2}
\end{figure*}

\begin{figure*}
\includegraphics[width=0.34\textwidth]{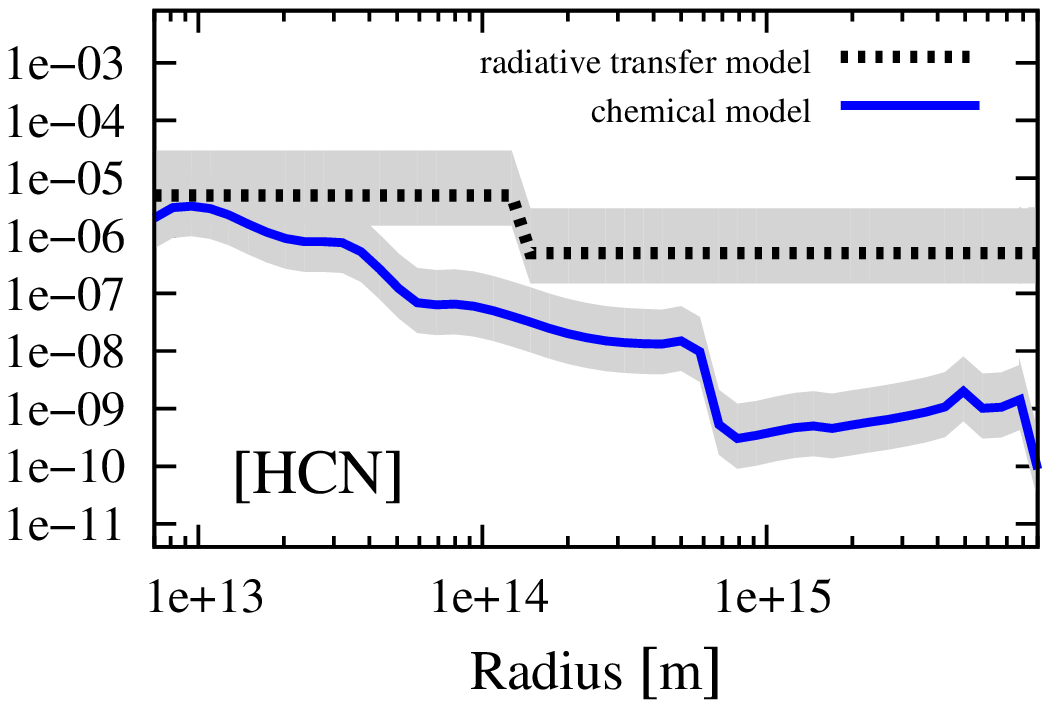}
\includegraphics[width=0.34\textwidth]{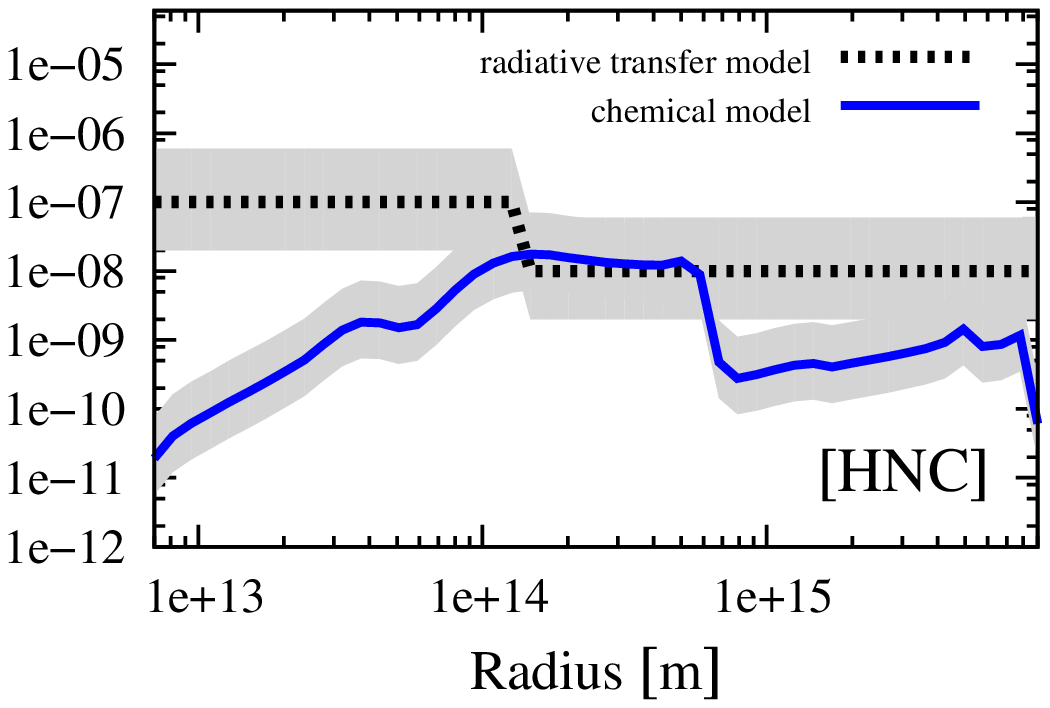}
\caption[]{The same as Fig. 2. Radiative transfer models are with a jump at 230\,K (higher inner abundance)
.} \label{fig:comp3}
\end{figure*}

\begin{figure*}
\includegraphics[width=0.34\textwidth]{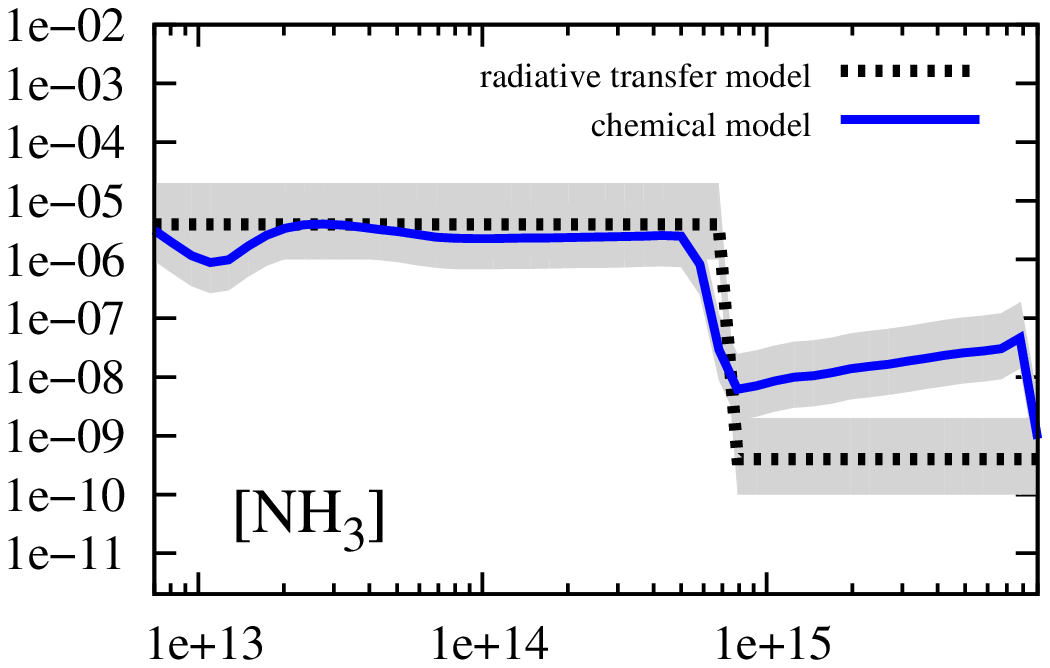}
\includegraphics[width=0.34\textwidth]{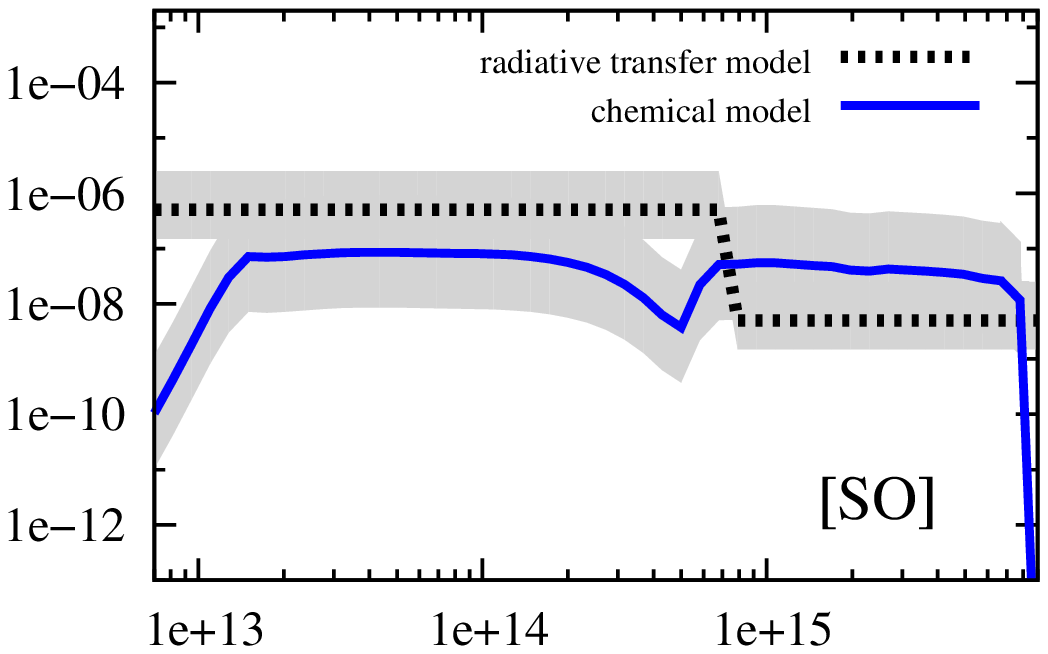}
\includegraphics[width=0.34\textwidth]{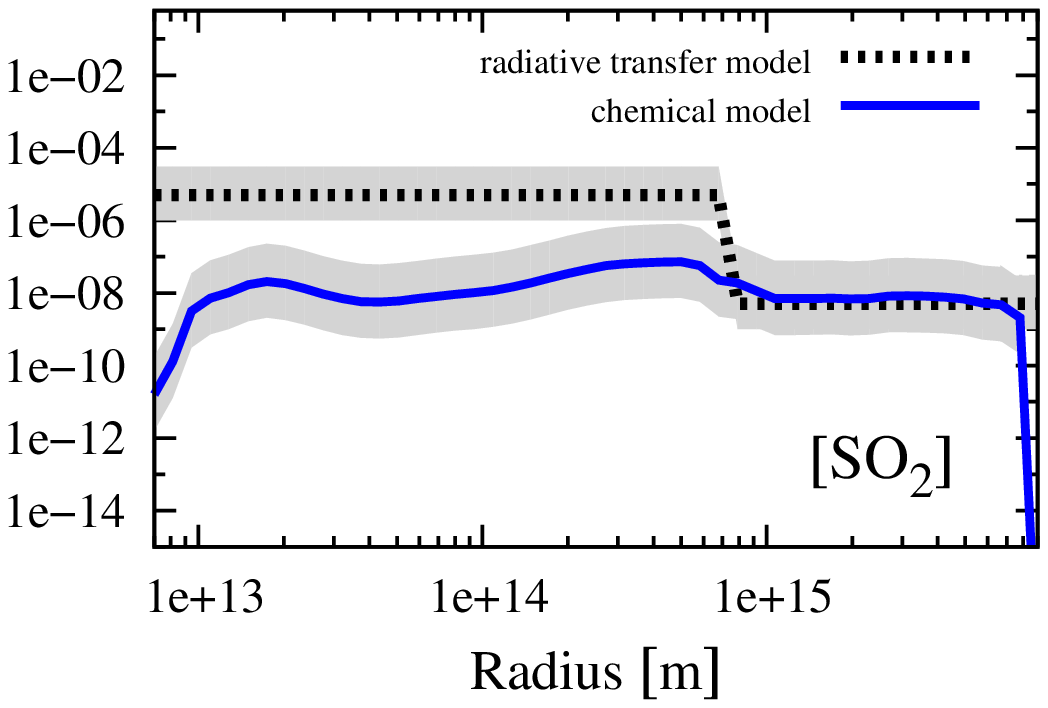}\\
\includegraphics[width=0.34\textwidth]{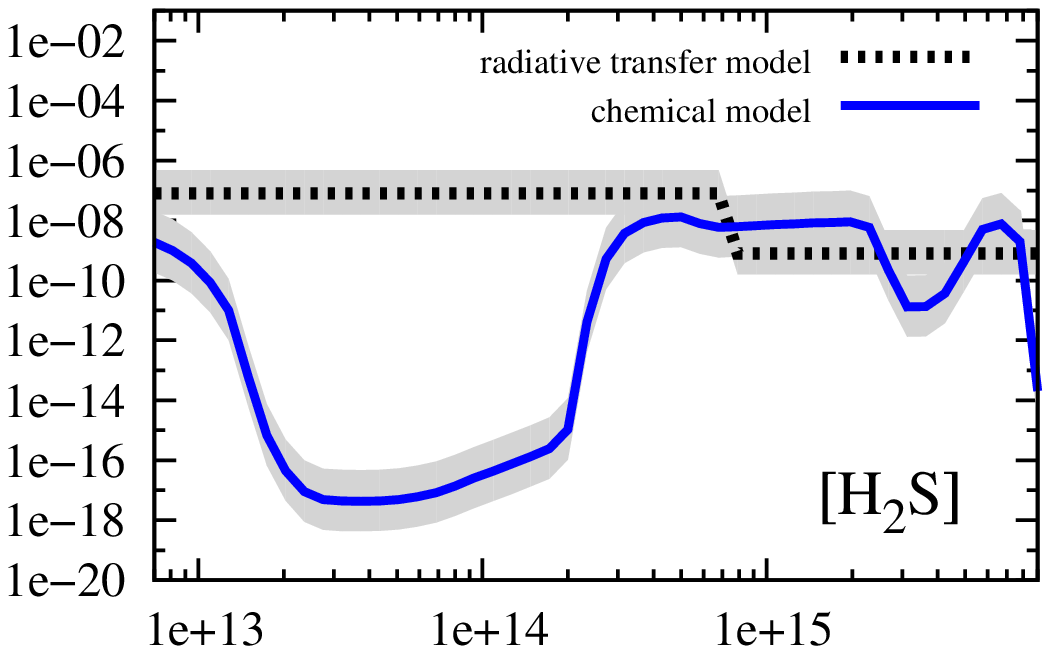}
\includegraphics[width=0.34\textwidth]{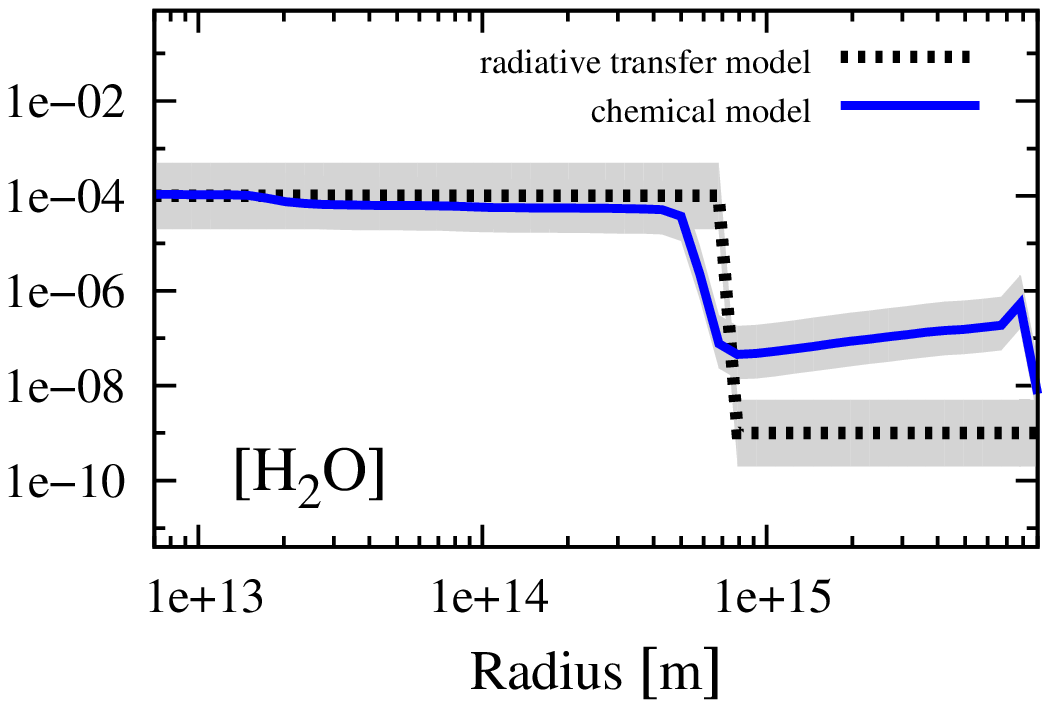}
\includegraphics[width=0.34\textwidth]{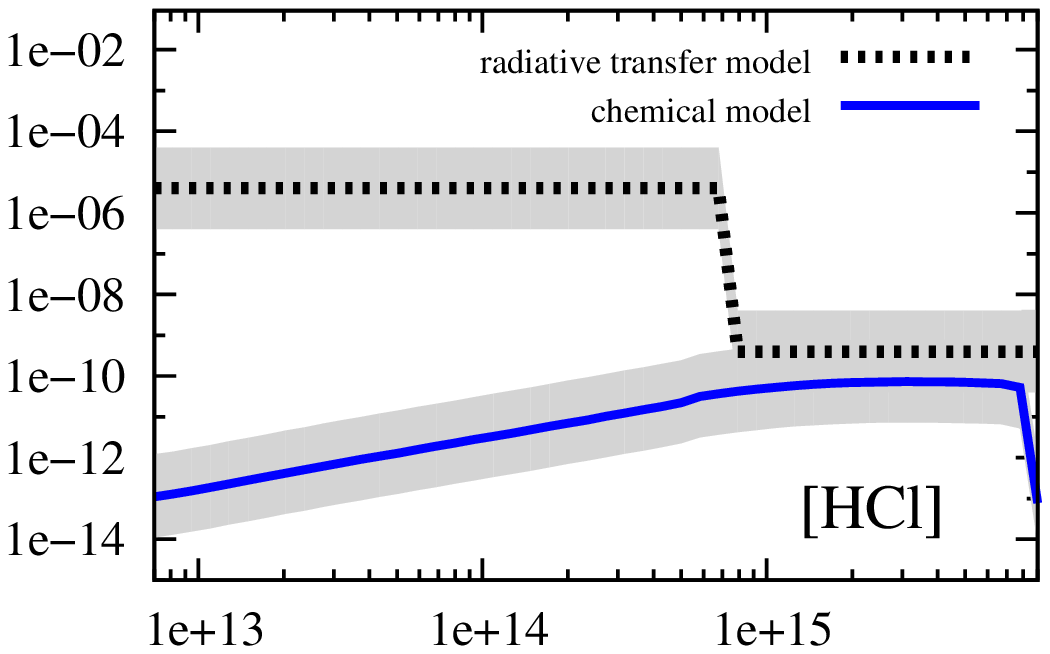}\\
\includegraphics[width=0.34\textwidth]{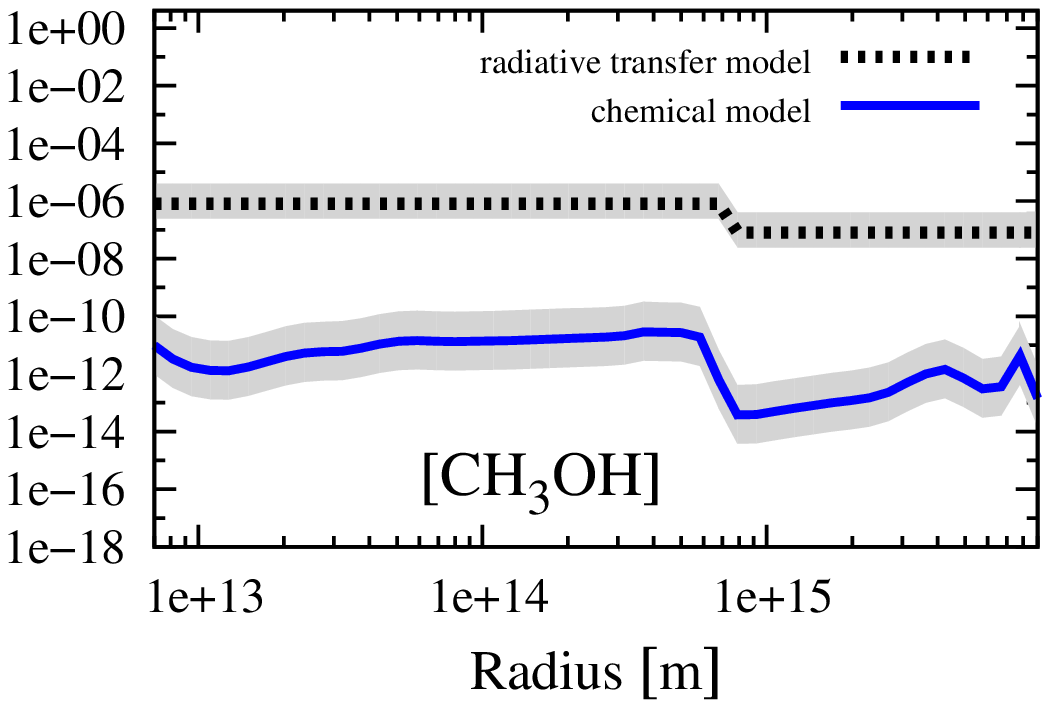}
\caption[]{The same as Fig. 2. Radiative transfer models are with a jump at 100\,K (higher inner abundance)
.} \label{fig:comp4}
\end{figure*}

%
\section{Chemical structure of the protostellar envelope} \label{Comparison}
%
%
With the 1D physical structure and chemical model described above, we calculated about 20 different realizations of the AFGL~2591 
envelope by varying the initial elemental abundances, chemical ages, grain sizes, UV intensities, and cosmic ray ionization rates. 

We identified several models that fit the \textsc{ratran}-based observational 1D abundance profiles reasonably well. Our best-fit 
model is based on the \textsc{lm} initial molecular abundances (with solar C/O ratio of 0.44; see 
Table~\ref{tab:init_abunds_afgl}), has a CR ionization rate of $5\times 10^{-17}$~s$^{-1}$, an external UV intensity of one Draine 
interstellar UV field, and a chemical age of $10-50$~kyr. The~model is compared with a \textsc{ratran}-derived observed values 
in 
\mbox{Figs.~\ref{fig:comp1}-\ref{fig:comp4}.} The~overall agreement between the observed and best-fit 1D abundance profiles is 
summarized in Table~\ref{tab:best_fit}.

As one can clearly see from Table~\ref{tab:best_fit}, more than $70\%$ of the 1D~\textsc{ratran} abundance profiles are reproduced 
within the uncertainties. The best fitted species are ortho- and para-H$_2$, CO, HCO$^+$, H$_2$CO, CN, N$_2$H$^+$, NO, OH, OCS, 
H$_2$CS, O, NH$_3$, SO, H$_2$O, and HNC. Species whose abundances in the 1D model are similar to the observationally derived 
values at some radii, but overall agreement is less clear are C, C$^+$, CH, CCH, CS, SO$_2$, H$_2$S, and HCl 
(Figs.~\ref{fig:comp1}-\ref{fig:comp4}). Species whose 1D abundances are not fitted include HCN and CH$_3$OH. Remarkably, many 
simplistic constant or jump-like 1D abundance profiles utilized in the \textsc{ratran} modeling of the observed lines correspond 
closely to the results of the rigorous 1D time-dependent chemical model.

%
\subsection{Photostable species located in the outflow cavity} \label{Comparison1}
%
%
Where the modeled abundance profiles for CCH, C, C$^+$, CH, CS, SO$_2$, and HCl coincide with the respective 
\textsc{ratran} profiles at some radii, the HCN and CH$_3$OH abundances in the best-fit model are lower than the 
\textsc{ratran}-derived values. The first reason for this disagreement is a fundamental restriction of the utilized 1D approach 
with which one cannot properly capture the physical and chemical processes that occur on the walls of UV-irradiated outflow 
cavity. The presence of the outflow cavity and even a disk-like rotating structure in the AFGL~2591 is well established (see, e.g., 
\citet{1992ApJ...391..710T}, \citet{Bruderer2010b}, and \citet{Wang2012}). Clearly, in this case such a complex environment is 
better characterized by a 2D or even 3D model.

As shown by \citet{Bruderer2010b}, a multidimensional approach is required to explain the observed abundances of light 
hydrides and hydride ions, such as OH$^+$. As we do not include a central UV source in our model (since its radiation in the 1D 
approach is quickly absorbed by the massive envelope inner regions), our model tends to under-predict the abundances of simple 
photostable radicals and PDR-like species. This may explain why the abundances of species like C, C$^+$, CH, CN, CCH, and HCl at 
small radii $(\la 10^{14}$~m) are lower in the model  than the \textsc{ratran}-derived observed values. For HCl, which has a very 
limited chemistry network in our model of only 58 channels, the lack of major high-temperature production channels may also play a 
role.

We increased the external UV intensity by factors of 50 and 500 compared to the best-fit model, but found that is has no effect 
on the calculated abundances  because the enhanced interstellar UV field cannot penetrate deeply into the massive 
AFGL~2591 envelope and the resulting UV-irradiated volume is  negligible in the 1D spherically-symmetric case, unlike in the 2D 
model with the outflow \citep[e.g., as described in][]{Bruderer2010b}. 

Thus, the 1D approach that is often employed to model and understand the chemical composition of low- and high-mass protostellar 
objects appears inappropriate for PDR-like ions and molecules. To explain their elevated abundances, a 2D or 3D model with a 
detailed description of the outflow physics (central X-ray or FUV source irradiating outflow cavity walls, outflow-induced shocks at 
the walls, etc.) must be utilized.

%
\subsection{HCN and HNC} \label{Comparison3}
%
%
Our models are able to fit either the HCN or the HNC abundance, but not both. This discrepancy reflects the limitations of 
modern astrochemical models that are not capable of explaining why the observed HNC/HCN abundance ratio decreases with increasing 
temperature in regions of low- and high-mass star formation (see recent discussion in \citet{2014ApJ...787...74G}). 
One explanation is that the energy barrier of the isomerization reaction converting HNC to HCN by 
collisions with atomic H, {\new quantum-chemically calculated to be $\sim$1200~K, should in fact  be lower, $\sim$200\,K}. 
Alternatively, a high-temperature destruction channel of HNC may be missing in the models,
{\new and one option is the reaction of HNC with O. The KIDA database gives a rate constant of \mbox{$\sim$10$^{-12}$\,cm$^3$\,s$^{-1}$} and a barrier of $>$1000\,K, which would suggest that this reaction only plays a minor role in the AFGL 2591 envelope. However, the reaction data are uncertain and we recommend an accurate measurement of this potentially important reaction}.
Third, the dynamical and thermal evolution of the environment may play a significant role in the HCN/HNC isomer-specific formation and 
destruction chemistry;   our model does not consider this effect as it is based on a static physical structure of the AFGL~2591 
envelope. 

%
\subsection{Complex organics} \label{Comparison4}
%
%
The problem with predicting CH$_3$OH abundances is that this species is produced as ice on dust grain surfaces during the cold, 
\mbox{pre-AFGL~2591} phase and is later injected into  the gas phase as a result of heating. 
Because our 1D model does not include a warm-up phase,
but rather  a simple two-stage cold-to-warm jump-like evolution, it probably does not allow methanol 
precursor ices to be sufficiently accumulated in grain mantles \citep[see][]{2008ApJ...682..283G}. In addition, we model the 
\mbox{pre-AFGL} phase assuming 0D physical conditions, which may not be sufficient for the 1D model used after that. It might also 
be that our approach to model CO hydrogenation on the dust surfaces does not include all the necessary details such as a multi-phase ice 
structure (chemically inert, bulk ice with chemically active ice layers), porosity, and the presence of surface binding sites of 
different energies (see \citet{2012A&A...538A..42T}, \citet{2013ApJ...762...86V,2013ApJ...769...34V}, and 
\citet{2013ApJ...778..158G}). This 
deficiency can be best demonstrated by the recent puzzling detections of complex organic molecules in cold, pre-stellar cores and 
infrared dark clouds, which is a challenge to explain without too much tuning of the key chemistry parameters such as reactive 
desorption efficiency, probabilities of radiative association between large molecules, and their least-destructive dissociative 
recombination \citep[e.g.,][]{2012A&A...541L..12B,2013ApJ...769...34V,2014ApJ...780...85V,2014ApJ...788...68O}. Interestingly, 
with our other models with elemental C/O ratios $>1$ we can accurately reproduce the entire 1D \textsc{ratran} CH$_3$OH profile, but 
then the H$_2$CO abundances in these models become too high. 

%
\subsection{S-bearing species} \label{Comparison5}
%
%
Another problem with many contemporary astrochemical models is simultaneously explaining the observed abundances of simple S-bearing 
species like CS and more complex species such as SO, SO$_2$, and OCS, and predicting the main reservoir of sulfur. The problem is 
particularly acute in protoplanetary disk chemistry \citep{2011A&A...535A.104D} and high-mass star-forming regions 
\citep[e.g.,~][]{2011A&A...529A.112W}. The reasons are a lack of data on sulfur chemistry from laboratory experiments and 
quantum chemistry simulations \citep{2010SSRv..156...13W}, and omission of key S-bearing complex species in the networks 
\citep[see][]{2012MNRAS.426..354D}. Another explanation is the lack of X-ray-driven chemistry in our model, which can be important 
for the evolution of S-bearing species in high-mass star-forming regions \citep[see~][]{Stauber2005,2007A&A...475..549B}.

{\bbf Recently, \citet{2014arXiv1406.2278E} have performed extensive chemo-dynamical modeling of the sulfur chemistry in 
Orion~KL with the UCL$\_$CHEM chemical code and two different evolutionary phases. In Phase~I, depletion of atoms 
and molecules onto dust grain surfaces begins in a collapsing prestellar core. In Phase~II, a protostar is formed and 
steadily heats the surrounding medium, evaporating the ices back into the gas phase. With this time-dependent physical and 
chemical model, \citet{2014arXiv1406.2278E} were able to simultaneously reproduce the observed column densities of  SO, SO$_2$, 
CS, OCS, H$_2$S, and H$_2$CS, assuming an initial S abundance of 10$\%$ of the solar value and a density $\ga 5\times 
10^{6}$~cm$^{-3}$.}

{\new In our models, only 1\% of cosmic sulfur is available for gaseous and surface chemistry, while the rest is locked in solid form.
In the standard {\sc lm} model, the main reservoirs of gas-phase sulfur are SO and SO$_2$, with H$_2$CS an important reservoir 
only at $T >$700\,K. In the other two models with C/O$>$1, CS-bearing species, especially CS and H$_2$CS, become the dominant
S-bearing species, while the importance of SO and SO$_2$ decreases.}

Our best-fit chemical model does reproduce reasonably well all the observed and analyzed S-bearing molecules in 
AFGL~2591 (CS, OCS, SO, SO$_2$, H$_2$CS, and H$_2$S; \mbox{see Figs.~\ref{fig:comp1}-\ref{fig:comp4}).} 
{\bbf Only SO$_2$ and 
H$_2$S are not well fitted in the inner region at $r\la 5\times10^{14}$~m$^{-2}$ (Fig.~\ref{fig:comp3}). The 1D modeled 
profile is almost constant for SO$_2$ and does not show a jump-like decrease at~$T=100$~K, as is found with the detailed 
\textsc{ratran} modeling. Still, the difference between the values calculated with the chemical and  \textsc{ratran} models is 
almost within the computational uncertainties.} 
{\new The situation is more severe for H$_2$S, where the 1D chemical model shows an abundance decrease of almost 8 orders 
of magnitude at $r \sim 3\times 10^{14}$~m, where neutral-neutral reactions with O and OH start destroying H$_2$S at $T>$140--200\,K. 
This jump-like behavior is not seen in the {\sc ratran} models, which may be due to our neglect of enhanced UV irradiation of the outflow cavity walls and 
inner envelope regions, 
which facilitate the formation of rich, heavy ices and their desorption into the gas phase.} 

%
\subsection{Other good models} \label{sec:other_best-fits}
%
In addition to our best-fit \textsc{lm} model with an age of $10-50$~kyr and a solar elemental C/O ratio of 0.4,  
we identified two other good 
models with \textsc{h11} initial abundances with an elemental C/O ratio of 1.13 (see~Table~\ref{tab:init_abunds_afgl}) and 
ages of $10$ and $50$~kyr. Unlike the standard model, with these two models the agreement is much better 
between the chemical and \textsc{ratran}-derived abundances of CH$_3$OH, HCN, HNC, CH, and CCH. However, these two models fail to 
reproduce the abundances of important species such as CO, HCO$^+$, H$_2$CO, and CN.

{\new The first reason is that the initial molecular abundances in the two C/O$>$1 models are calculated assuming a warmer $T=15$~K than the $T=10$~K in the 
\textsc{lm} model. 
The higher temperature leads to faster diffusion rates for atomic hydrogen and
hence more efficient hydrogenation of CO ice into H$_2$CO and CH$_3$OH ice.
As a result, the abundances of gaseous H$_2$CO and CH$_3$OH become higher
and are in better agreement with the observed values than with the standard {\sc lm} model.}

The second reason is the excess 
of elemental carbon that remains available after almost all elemental oxygen is locked in CO and CO$_2$. This carbon goes into 
production of  various 
hydrocarbons and cyanopolyynes (Table~\ref{tab:init_abunds_afgl}), making HCN, HNC, CH, and CCH more abundant and better fitted 
than with the \textsc{lm} solar C/O model.

%
\subsection{Influence of different parameters on chemical abundances} \label{influence}
%
The overall sensitivity of the modeled molecules to variations of the physical and chemical parameters of the 1D model is 
summarized in Table~\ref{tab:species_sens}. 

First, we  find that the modeling results are not sensitive to the adopted value of the average grain size. This is not 
surprising because the envelope is warmer than $30$~K and has a rather low age of around 1 $-$ 5 $\times ~10^4$~years. Under 
such conditions the ices that were produced in the earlier, colder \mbox{pre-AFGL~2591} phase have mostly evaporated and surface 
processing becomes of minor importance for the global chemical evolution of the envelope.

The same is true for two other models where the external UV intensity was 50 and 500 (in Draine's interstellar UV units) 
compared to the best-fit model with 1 Draine UV field. As we stated above, this is because the external UV field cannot penetrate 
deeply into the massive, dense AFGL~2591 envelope, and the corresponding propagation of the PDR-like outer shell inward is 
negligible.

%
\subsubsection{Ionization rate} \label{Ionization_rate}
%
%
The species that are sensitive to the enhanced CR ionization rate values of \mbox{$5\times10^{-16}$~s$^{-1}$} and 
\mbox{$5\times10^{-15}$~s$^{-1}$} include H$_2$O, molecular ions like HCO$^+$ or N$_2$H$^+$, and organics such as 
H$_2$CO and H$_2$CS. Because of  the higher ionization in these models, the modeled abundances of ions are over-predicted compared to the 
best-fit model and the \textsc{ratran}-derived values, whereas abundances of water and the organic species are under-predicted. 
The rest of the considered molecules are not very sensitive to the adopted CR ionization rate.

%
\subsubsection{C/O ratio} \label{CO_ratio}
%
%
Most of the species in Table~\ref{tab:species_sens} are sensitive to the adopted initial abundances and the C/O ratio. The only 
exceptions are several simple H- and N-bearing species (H$_2$, N$_2$H$^+$), C and C$^+$, and HCl. Both C and C$^+$ are 
abundant in the outer, UV-irradiated part of the envelope, where they are the dominant carbon-based species. Consequently, their 
abundances differ by only a factor of about 2 between the \textsc{lm} and \textsc{d11c} and \textsc{h11} models 
(see~Table~\ref{tab:init_abunds_pre}). 
The abundances of HCl, and also of H$_2$ and N$_2$H$^+$, are not sensitive to the C/O ratio because their major 
formation and destruction pathways do not include reactions with C- or O-bearing species. 

%
\subsubsection{Chemical age} \label{Chemical_age}
%
%
Molecules whose 1D abundance profiles are affected by the chemical age are mostly 
heavy S-bearing and N-bearing species, namely, H$_2$CS, OCS, NH$_3$, H$_2$S; and HCN  and HNC 
(Table~\ref{tab:species_sens}). Less sensitive to the chemical age are the abundances of H$_2$CO, CH, and CS. As
established more than two decades ago by e.g.,~\citet{1992ApJ...392..551S}, abundance ratios of sulfur- and nitrogen-bearing species such as 
CCS / NH$_3$ can be employed as chemical age estimators in cold pre-stellar cores, because  
sulfur-bearing species like CCS are formed rather rapidly via gas-phase chemistry involving hydrocarbons (CH, CCH, etc.), 
whereas nitrogen chemistry takes longer to convert N to N$_2$ and to other N-bearing species like HCN and HNC. Naturally, 
characteristic chemical timescales of NH$_3$, which is mainly formed on dust grain surfaces, are  rather long ($>10^5$~years 
in molecular clouds). This is also 
true for other complex species produced via surface chemistry, like H$_2$CO, H$_2$CS, and CH$_3$OH. As we discussed above, the 
best-fit models for the AFGL~2591 1D chemical structure are those with short chemical ages of $10-50$~kyr, which is similar to the 
derived age of the AFGL~2591 of $\sim 50$~kyr \citep{Doty2002}.

%
\section{Conclusions} \label{Conclusions}
%
%
Based on HIFI and JCMT observations we derived molecular abundances of about 25 species detected in the protostellar envelope of 
AFGL~2591. Two methods were applied to interpret the data. First, detailed radiative transfer calculations with 
\textsc{ratran} were performed, using an adequate 1D physical model of the envelope. Then, abundances were calculated with 
time-dependent chemical models with varied cosmic ray ionization rates, external UV intensities, grain sizes, ages, and initial 
molecular abundances based on various C/O elemental abundances. 
By comparing the results from these two methods, we reach the following conclusions:
\begin{enumerate}
\item The radial abundance profiles for 25 species were derived; i.e., H$_2$, CO, CN, CS, HCO$^+$, H$_2$CO, N$_2$H$^+$, CCH, NO, 
OCS, OH, H$_2$CS, O, C, C$^+$, CH, HCN, HNC, NH$_3$, SO, SO$_2$, H$_2$S, H$_2$O, HCl, and CH$_3$OH.
\item Based on radiative transfer calculations, three molecules (CO, CN, CS) were firmly fitted by a constant abundance model. The 
line fluxes of HCO$^+$, H$_2$CO, N$_2$H$^+$, and CCH can be fitted equally well by a constant abundance model and by a model with a 
jump at 100\,K, with lower inner abundances. Species such as NO, OCS, OH, H$_2$CS, O, C, C$^+$, and CH were fitted by a constant 
abundance model; however, because of a lack of data, models with jumps cannot be excluded. The best model for HCN and HNC was with a 
jump at 230\,K, with a higher inner abundance. The best model for molecules such as NH$_3$, SO, SO$_2$, H$_2$S, H$_2$O, HCl, and 
CH$_3$OH was with a jump at 100\,K, again with a higher inner abundance.
\item The best time-dependent chemical model predicts a chemical age of AFGL~2591 {\bbf of $\sim 10-50$~kyr}, a solar C/O 
ratio of $0.44$, and a cosmic ray ionization rate of $\sim 5  \times 10^{-17}$~s$^{-1}$.
\item {\new Among our considered 25 species, the time-dependent chemical model can explain the abundance profiles of 15 of them within the errors}. The remaining molecular abundances are affected by limited reaction networks (Cl- and S-bearing species), the lack of observed transitions (e.g., C, O, C$^+$, and CH), or deficiencies of the surface chemistry model (CH$_3$OH), which cause higher uncertainties and reduced agreement.
\item We found that grain properties and intensity of the external UV field do not strongly affect the chemical structure of the 
AFGL~2591 envelope (at least with the adopted 1D geometry), whereas the chemical age, cosmic ray ionization rate, and initial 
abundances play an important role.
\item Molecules that are sensitive to the presence of the outflow and UV/X-ray irradiation are C, C$^+$, CH, CCH, CS, SO$_2$, and other S-bearing molecules.
\item Molecules that are sensitive to the cosmic ray ionization rate are  H$_2$O, molecular ions like HCO$^+$, N$_2$H$^+$, and organic species such as H$_2$CO and H$_2$CS.
\item Molecules that are sensitive to the adopted chemical age include heavy H$_2$CS, OCS, NH$_3$, H$_2$S;  HCN and HNC; and to a lesser degree H$_2$CO, CH, and CS.
\item The only molecules that are not sensitive to the initial molecular abundances and C/O ratios are simple H- and N-bearing species like H$_2$, N$_2$H$^+$; C and C$^+$; and HCl.
\end{enumerate}

The main conclusion of this work is that simple constant or jump-like abundance profiles reproduce accurately single-dish submillimeter observations, and are supported by the predictions of a time-dependent gas-grain chemical model bound to a 1D physical structure. Future studies aimed at explaining the chemical structure of all observed species in the AFGL~2591 envelope should consider 2D or even 3D envelope geometry with an outflow cavity, and include the evolution of its temperature and density and, possibly, X-ray irradiation from the central source(s).

%
\begin{acknowledgements}
%
%
We thank Robert Barthel for his help with the calculations, Veronica Allen for useful suggestions, {\bbf and an anonymous referee for a helpful report}.
\\
D. Semenov acknowledges support by the {\it Deutsche Forschungsgemeinschaft} through SPP~1385: ``The first ten million years of the solar system - a planetary materials approach'' (SE 1962/1-3). \\
HIFI has been designed and built by a consortium of institutes and university departments from across Europe, Canada and the United States under the leadership of SRON Netherlands Institute for Space Research, Groningen, The Netherlands and with major contributions from Germany, France and the US. Consortium members are: Canada: CSA, U.Waterloo; France: CESR, LAB, LERMA, IRAM; Germany: KOSMA, MPIfR, MPS; Ireland, NUI Maynooth; Italy: ASI, IFSI-INAF, Osservatorio Astrofisico di Arcetri-INAF; Netherlands: SRON, TUD; Poland: CAMK, CBK; Spain: Observatorio Astron{\'o}mico Nacional (IGN), Centro de Astrobiología (CSIC-INTA). Sweden: Chalmers University of Technology - MC2, RSS \& GARD; Onsala Space Observatory; Swedish National Space Board, Stockholm University - Stockholm Observatory; Switzerland: ETH Zurich, FHNW; USA: Caltech, JPL, NHSC. \\
This research made use of NASA's Astrophysics Data System.
\end{acknowledgements}

%
%

\bibliographystyle{aa}
\bibliography{aa24657}

\end{document}